\documentclass[usenatbib]{mnras}

\usepackage{graphicx}
\usepackage{txfonts}
\usepackage{enumerate}
\usepackage{xcolor} 
\usepackage{aas_macros} 
\newcommand{\MAA}{\textsc{\AA}}

\title[The evolution of high-z galaxy colours]{Interpreting the evolution of galaxy colours from $z = 8$ to $z = 5$}
\author[Mancini et al.]{Mattia Mancini$^{1,2,3}$\thanks{E-mail:
mattia.mancini@oa-roma.inaf.it}, Raffaella Schneider$^{1,3}$, Luca Graziani$^{1,3}$, Rosa Valiante$^{1}$,
\newauthor 
Pratika Dayal$^{4}$, Umberto Maio$^{5,6}$, Benedetta Ciardi$^{7}$\\
$^{1}$INAF/Osservatorio Astronomico di Roma, Via di Frascati 33, 00040 Monte Porzio Catone, Italy\\
$^{2}$Dipartimento di Fisica, ``Sapienza'' Universit{\'a} di Roma, Piazzale Aldo Moro 5, 00185, Roma, Italy\\
$^{3}$Kavli Insitute for Theoretical Physics, University of California Santa Barbara, CA 93106 \\
$^{4}$Kapteyn Astronomical Institute, University of Groningen, P.O. Box 800, 9700 AV Groningen, The Netherlands   \\
$^{5}$ INAF - Osservatorio Astronomico di Trieste, via G. B. Tiepolo 11, 34131 Trieste, Italy \\
$^{6}$Leibniz Institute for Astrophysics, an der Sternwarte 16, 14482 Potsdam, Germany\\
$^{7}$Max Planck Institut f¨ur Astrophysik, Karl-Schwarzschild-Strasse 1, 85741 Garching, Germany   
}
\begin{document}
\date{}

\pagerange{\pageref{firstpage}--\pageref{lastpage}} \pubyear{2016}

\maketitle

\label{firstpage}

\begin{abstract}
We attempt to interpret existing data on the evolution of the UV luminosity function and UV colours, $\beta$, of galaxies at $5 \leq z \leq 8$, to improve our understanding of their dust content and ISM properties. To this aim, we post-process the results of a cosmological hydrodynamical simulation with a chemical evolution model, which includes dust formation by supernovae and intermediate mass stars, dust destruction in supernova shocks, and grain growth by accretion of gas-phase elements in dense gas. We find that observations require a steep, Small Magellanic Cloud-like extinction curve and a clumpy dust distribution, where stellar populations younger than 15 Myr are still embedded in their dusty natal clouds. Investigating the scatter in the colour distribution and stellar mass, we find that the observed trends can be explained by the 
presence of two populations: younger, less massive galaxies where dust enrichment is mainly due to stellar sources, and massive, 
more chemically evolved ones, where efficient grain growth provides the dominant contribution to the total dust mass. 
Computing the IR-excess - UV color relation we find that all but the dustiest model galaxies follow a relation shallower than the 
Meurer et al. (1999) one, usually adopted to correct the observed UV luminosities of high-$z$ galaxies for the effects of dust extinction. 
As a result, their total star formation rates might have been over-estimated. Our study illustrates the importance to incorporate a proper treatment of dust in simulations of high-$z$ galaxies, and that massive, 
dusty, UV-faint galaxies might have already appeared at $z \lesssim 7$.
\end{abstract}

\begin{keywords}
dust, extinction – ISM: supernova remnants – submillimetre: galaxies – galaxies:
evolution – galaxies: high-redshift – galaxies: ISM
\end{keywords}

\section{Introduction}

In the last decade, data from the {\it Hubble Space Telescope}\footnote{http://www.stsci.edu} (HST), especially after the advent of the Wide Field Camera (WFC3), 
allowed us to collect large samples of galaxies at $z \sim 7 - 8$, with smaller samples extending up to $z \sim 9 -11$ \citep{McLure2013, Bouwens2014, Oesch2014, Bouwens2015a, McLeod2015, Finkelstein2015}, 
among which the two most distant spectroscopically confirmed galaxies at $z = 8.68$ \citep{Zitrin2015} and $z = 11.1$ \citep{Oesch2016}.
Since spectroscopic observations of galaxies at $z>6$ with ground-based telescopes are very challenging, observers have developed alternative, photometry-based 
techniques to both select high-$z$ candidates and estimate their physical properties. 
For example, the total stellar mass, the stellar age and the ongoing star formation rate (SFR) can be estimated from spectral energy distribution (SED) fitting and colour index analyses.

Two key quantities are generally used to characterize the properties of the first galaxies and of their dominant stellar populations: the UV luminosity function (LF) and the 
observed UV spectral slope, $\beta$ ($f_\lambda \propto \lambda^\beta$, \citealt{meurer1999}). The LF, defined as the number density of galaxies per unit magnitude, provides important constraints on star formation efficiencies at different 
redshifts and on their evolutionary status, especially at early times \citep{Bouwens2011}.  As the first structures collapse and assemble their stellar content, the inter-stellar medium (ISM) 
is progressively enriched with metals and dust. Dust extinction affects the UV luminosity and should leave a signature in the galaxy LF.  While for low-redshift galaxies dust extinction can be corrected by 
measuring the far infrared emission (FIR), observations of high-$z$ galaxies with millimeter (mm) telescopes, such as the {\it Atacama Large Millimeter Array} (ALMA)
and the {\it Plateau de Bure Interferometer} (PdBI), have mostly provided
upper limits on the rest-frame FIR emission of $z > 6$ UV-selected galaxies  \citep{Kanekar2013, Ouchi2013, Ota2014, Schaerer2015, Maiolino2015, Zavala2015}, with one notable exception \citep{Watson2015, Knudsen2016}. 
For this reason, it has become a common practice to estimate the effects of dust extinction using the observed $\beta$ slope, or UV colour.

Despite a vigorous debate in the past ten years, recent observational results appear to converge on a common trend for the shape and the evolution of the UV LF in the redshift range $4 < z < 8$ \citep{Bouwens2015a},
down to an AB magnitude of $-16$ at $z = 4, 5$ and of $\sim -17$ at $z = 6, 7$ and $8$. At even higher redshifts, even the deepest
observations in blank fields can only probe the bright-end of the LF, providing important constraints on the
volume density of the most luminous galaxies with $M_{\rm UV}< -20$ at $z = 9$ and $10$  \citep{Bouwens2015a}. An efficient way
to push the observations to fainter luminosities is to exploit the gravitational lensing magnification of massive galaxy clusters. Results
from the HST programs CLASH and Hubble Frontier Fields (HFF) have increased the statistics of candidate galaxies at the highest redshifts,
providing better constraints on the evolution of the faint-end slope of the LF and placing the first limits on the LF at $z \sim 10$ \citep{Atek2015, 
McLeod2015, McLeod2016, Livermore2016} .
 
It is custumary to fit the LF with a Schechter function, which has a power-law
behaviour with slope $\alpha$ at the faint-end, an exponential cut-off brighter than a characteristic luminosity (magnitude) $L_\ast$ ($M_\ast$) and a volume density of $\phi_\ast$ at this
characteristic luminosity,
\begin{equation}
\frac{dn}{dL} = \phi(L) = \left(\frac{\phi_\ast}{L_\ast}\right)\left(\frac{L}{L_\ast}\right)^\alpha e^{-L/L_\ast}.
\end{equation} 
In general, the evolution of the LF with redshift is characterized by means of variations of these Schechter parameters and is consistent with a steady growth in the volume
density and luminosity of galaxies with time. In particular, there is a significant evidence for a steepening of the faint-end slope with $z$, 
in agreement with the predicted steepening of the halo mass function, a modest evolution of $M_\ast$ and a decrease of $\phi_\ast$ from $z \sim 4$ to $z \sim 7$
 \citep{Bouwens2015a}.  Some observations at $z = 9$ and $10$ suggest a faster evolution, and that the luminosity densities inferred from current samples 
are $\sim 2$ times lower than the values extrapolated from the trends at $4 < z < 8$ \citep{Bouwens2015a}. Other studies support a smoother evolution
from $z = 8$ to 9 \citep{McLeod2015, McLeod2016, Finkelstein2015}. Indeed, the recent discovery of GN-z11, a luminous galaxy with $\rm M_{UV} = - 22.1$ at $z = 11$ \citep{Oesch2016}
may indicate that the LF at the very bright end does not follow a Schechter functional form, 
possibly due to less efficient feedback at very high redshifts \citep{Bowler2014, Dayal2014,Finkelstein2015b,Waters2016}. 

To convert the observed UV luminosity to a SFR and compare the above findings to theoretical predictions, dust correction is usually estimated using the observed $\beta$ slopes and the so-called IRX-$\beta$
relationship by \citet{meurer1999}, who proved that, at $z <3$, the amount of SED reddening directly correlates with the $\beta$ value, as also confirmed by independent theoretical predictions (e.g. \citealt{Wilkins2012b}). 
Although this relation has been calibrated on starburst galaxies at low redshifts, and assumes a constant mean intrinsic slope of $\beta = -2.23$, this procedure has been widely adopted in high-$z$
galaxy surveys \citep[see][]{Bouwens2012}. However, the value of $\beta$ is also a function of important properties of the stellar populations, such as their ages, metallicity and initial mass function (IMF). 
Although with large uncertainties, observational trends have been reported which quantify the dependence of $\beta$ on the UV luminosity and redshift  \citep{Stanway2005, Wilkins2011, Finkelstein2012, Bouwens2012, Castellano2012, Dunlop2013}.
In general, the observations are consistent with a decreasing reddening towards lower luminosities and higher redshift. A coherent analysis of the observed $\beta$ for galaxies in a wide redshift range, from $z \sim 4$ to $z \sim 7$, has been recently made by \citet{Bouwens2014},
who confirm a strong evidence for a dependence of the average $\beta$ on the UV luminosity, the so-called Colour-Magnitude-Relation (CMR, \citealt{Rogers2014}), 
with brighter galaxies being redder and fainter galaxies being bluer, and a flattening of the relation at luminosities faintward of $M_{\rm UV} \sim -19$. They also report a small but
clear evolution with time, with galaxies at fixed luminosity becoming bluer with $z$. For the faint galaxies with $-19 < M_{\rm UV} < -17$, the mean $\beta$ at $z \sim 4, 5$ and $6$ is $-2.03$, $-2.14$ and $-2.24$ respectively. Extrapolation of this
trend to $z \sim 7$ and $8$ suggests mean values of $-2.35$ and $-2.45$, consistent - within the errors - with the observed ones. 

Theoretical studies have attempted to interpret the data by means of numerical simulations or semi-analytical models. 
\citet{Wilkins2012b} explored the sensitivity of the intrinsic $\beta$ slopes to the IMF and to the recent star formation and metal enrichment histories of low-$z$ galaxies.
They find a distribution of $\beta$ values with a scatter of $0.3$, which introduces an uncertainty in the inferred dust attenuation when a constant intrinsic slope is assumed. 
This scatter is significantly reduced for galaxies at $z \sim 6$, but the mean intrinsic $\beta$ decreases with $z$. If this is not properly taken into account and the locally calibrated relation is applied, dust
attenuation is systematically underestimated \citep{Wilkins2013}. \citet{GonzalezPerez2013} have demonstrated the dependence of the galaxy UV colours on the adopted dust properties and, in
particular, on the dust extinction curve. With the aim of intepreting high-$z$ Lyman-$\alpha$ emitters (LAEs) and Lyman Break Galaxies (LBGs) observations, \citet{Dayal2010a} and \citet{Dayal2012} used a numerical simulation to derive intrinsic galaxy properties and a semi-analytical model
to estimate dust attenuation. They explored the resulting UV LF and the dependence of $\beta$ on the galaxy UV luminosity with and without dust attenuation. 
They found that dust attenuation improves the agreement with the observations, but the observed CMR was not reproduced by  the model results. More recently, \citet{Khakhaleva2016} post-processed the results of cosmological simulations with a simple dust model that assumes a
constant dust-to-metal mass ratio in the neutral gas and that dust is instantaneously sublimated in hot ionized regions. They used a Monte Carlo radiative transfer code to predict UV attenuation and IR re-emission of their model galaxies.  By means of a detailed comparison with observations at $5 \le z \le 10$,
they concluded that, in order to assess the effects of dust in the ISM of high-$z$ galaxies, the complex interplay
of dust creation and destruction processes should be fully incorporated into numerical simulations. 

With this aim, in \citet{mancini2015} we have presented a semi-numerical model which includes a physically motivated description of dust evolution, accounting for dust enrichment by Supernovae (SNe) and 
Asymptotic Giant Branch (AGB) stars, the effects of dust destruction by SN shocks and grain growth in the dense cold phase of the ISM (see also \citealt{Valiante2009, Valiante2011, deBennassuti2014}). 
We then compared the model predictions with the limits on the dust mass inferred from mm-observations of $z > 6$ galaxies, deriving 
interesting constraints on the properties of their ISM and on the nature of dust at high-$z$. 
Here we extend this previous investigation with the goal of intepreting the observed UV luminosities and colours of galaxies at $5 < z < 8$.

The paper is organized as follows. In Section \ref{sec:method} we describe our method and the assumptions made to compute the dust content and 
luminous properties of the simulated galaxies. 
In Section \ref{sec:results} we first discuss the predicted physical properties of the galaxies at $5 \le z \le 8$. Then we derive the UV LFs and
$\beta$ slopes assuming no dust extinction, and discussing the dependence of the results on the extinction model. In Section \ref{sec:comparison}
we introduce the
model that better reproduce the observed UV LFs and CMR. We analyze the origin of the scatter around the mean values 
at different $z$, both in the CMR and in the stellar mass - UV luminosity relation. We compute the IR excess and dust attenuation factors, comparing
with observationally inferred correlations. Finally, in Section~\ref{sec:conclusions} we summarize the results and draw the main conclusions.

\section{Method}
\label{sec:method}
In this section we describe the semi-numerical model that we have developed. First, we infer the intrinsic galaxy properties using the output of a hydro-dynamical simulation of structure formation described in Section \ref{met:cosmo}. To compute the {\it intrinsic} UV luminosity of each galaxy, we calculate the spectral energy distribution (SED) as described in Section~\ref{met:spectra}.  
We then couple the simulated output, in particular the star formation rate (SFR), metallicity ($Z$) and mass of gas ($M_{\rm g}$) of each simulated galaxy, with a semi-analytical model to estimate the dust mass ($M_{\rm d}$) as in \citet{mancini2015}. A brief summary of this method is provided in Section \ref{met:chem}.  Finally, in Section \ref{met:extinction}, we present the method adopted to compute dust extinction.

\subsection{Cosmological simulation}
\label{met:cosmo}
We use the $\Lambda$CDM\footnote{The adopted cosmological parameters are  $\Omega_{\rm M}=0.3$, $\Omega_\Lambda=0.7$, $\Omega_{\rm b}=0.04$, $h = 0.7$, $n = 1$ and $\sigma_8 = 0.9$, consistent with
WMAP7 data release \citep{Komatsu2011}.} cosmological simulation presented in \citet{maio2010}. Here we briefly describe only the main features of the code and the set-up of the simulation;  
we refer the interested reader to the original paper for more details.

The simulation has been run using a modified version of the GADGET2 code \citep{springel2005}. 
It is able to predict the star formation rate and the metal production in each collapsed object, accounting for the evolution of hydrogen, 
helium and deuterium \citep{yoshida2003, maio2007}. The effect of resonant transition of molecules in the gas cooling function is also accounted for, as described in \citet{maio2007}.
Star formation and metal enrichment are computed taking into account stellar lifetimes using the lifetime function from \citet{padovani1993}, while the metallicity dependent metal yields adopted in the simulation are from  \citet{woosley1995} for core-collapse SNe, \citet{vandenhoek1997} for AGB  stars, \citet{thielemann2003} for type-Ia SNe  and from  \citet{heger2002} for pair-instability SNe.  At each given time, the stellar IMF depends on the metallicity of the star forming
gas as described in  \citet{tornatore2007}. In particular, we assume that Pop~III stars form according to a Salpeter IMF with masses in the range $100\, M_\odot \leq m \leq 500 \, M_\odot$ when the $Z < Z_{\rm cr} = 10^{-4} \, Z_\odot$.
Above this threshold, we assume that metal-fine structure cooling is efficient enough to trigger low-mass Pop~II/I star formation and the Salpeter IMF is shifted to the mass range $0.1\, M_\odot \leq m \leq 100 \, M_\odot$.
Feedback from SN explosions is taken into account by modelling a multi-phase interstellar medium  (ISM) as in  \citet{Springel2003} and successive developments presented in \citet{Maio2011}. 
A uniform, redshift-dependent UV background produced by quasar and galaxies is also assumed \citep{Haardt1996}.

The simulation box has a size of 30 $h^{-1}$ Mpc (comoving), with periodic boundary conditions. The total number of dark matter and gas particles is $N_{\rm p} = 2 \times 320^3$ and the dark matter (gas) particle mass is 
$M^{\rm p}_{\rm DM}=6 \times10^7 \, h^{-1} M_\odot$ ($M^{\rm p}_{\rm g}= 9 \times 10^6 h^{-1} M_\odot$).  In Sec.~\ref{res:physprop} we discuss the impact of the simulation volume and mass resolution on this study.
Dark matter halos are identified by means of an FOF algorithm as a group of at least 32 gravitationally bound particles.

By tracking the star particles along the redshift evolution we also reconstruct the merger tree of each simulated galaxy. This is needed to compute the evolution in redshift of the star formation rate, the gas mass and metallicity,
which have been used to initialize the dust evolution model, as detailed in Section \ref{met:chem}.

\subsection{Intrinsic galaxy spectra}
\label{met:spectra}

The {\it intrinsic} SED of a galaxy depends on the IMF, age and metallicity of each stellar population that contributes to the emission. 
Since the mass fraction of active Pop~III stars is negligible\footnote{We find that the mass fraction of Pop~III stars decreases with
the UV luminosity ranging from $\sim 0.1$ at $M_{\rm UV} = -18$ to $\sim 10^{-3}$ at $M_{\rm UV} = -22$. However, in all but 2 galaxies at
$z \sim 6$, Pop~III stars have already disappeared due to their short lifetimes, and their contribution to the UV emission at $z < 8$ can be safely neglected.} at $5 < z < 8$
(see also \citealt{Salvaterra2011}), we consider the UV luminosity contributed by Pop~II/I
stars using the spectral synthesis model StarBurst99 (hereafter SB99, \citealt{Leitherer1999, Vazquez2005}).
We assume that each star particle, which represents a single stellar population, is formed in an istantaneous burst.
The routines of SB99 responsible for computing dust extinction have been disabled and only the stellar and nebular emission  is accounted for. In this way, we 
compute a database of intrinsic spectra in the metallicity range $0.02 \, Z_\odot \leq Z_\ast  \leq 1 \,Z_\odot$ for stellar ages $2 \, {\rm Myr} \le t_{\star} \leq 1 \, {\rm Gyr}$. 
The database is used to assign a spectrum $f^i_\lambda (Z_\ast, t_\ast)$ to each star particle $i$ and the cumulative {\it intrinsic} SED of the $j$-th galaxy to which the star particle belongs 
is computed as,

\begin{equation}
F^j_\lambda = \sum_i f^i_\lambda(Z_\ast, t_\ast) \, M_{\ast,i}, 
\label{eq:int_sed}
\end{equation}
\noindent
where $M_{\ast,i}$ is the mass of the $i$-th star particle.

\subsection{Dust evolution model}
\label{met:chem}

Dust grains can form by condensation of gas-phase metals in the ejecta of SNe \citep{Todini2001,Nozawa2003,Schneider2004,Bianchi2007,Cherchneff2008,Cherchneff2009,Sarangi2013, Marassi2014,Marassi2015} 
and in the atmosphere of AGB stars \citep{Ferrarotti2006,Zhukovska2008,Ventura2012a,Ventura2012b,Dicriscienzo2013,Nanni2013,Ventura2014,Schneider2014}.
Once created by stars and dispersed in the interstellar medium of a galaxy, dust grains evolve depending on the environmental conditions. In the dense cold phase of the ISM, dust grains can grow by accretion of gas-phase 
elements \citep{Asano2013,Hirashita2014a} while in the hot diffuse phase the grains can be efficiently destroyed by interstellar shocks \citep{Bocchio2014}. All these processes are reviewed in \citet{Draine2011} and have been 
implemented in chemical evolution models with dust \citep{Valiante2009,Valiante2011,deBennassuti2014}. 
Since the cosmological simulation does not have the resolution to describe the different phases of the ISM which are relevant to dust evolution, 
the values it provides are only indicative of the average physical conditions of their ISM. To circumvent this limitation, we follow the same approach adopted in  \citet{mancini2015},
where dust enrichment in each galaxy can be described self-consistently within the average properties predicted by the simulation.

Following  \citet{deBennassuti2014}, we adopt a 2-phase ISM model with a diffuse component (warm/hot low-density gas), where dust can be destroyed by SN
shocks, and a dense or molecular cloud component (cold and dense gas), where star formation and grain growth occur.  
The time evolution of the ISM mass (gas and dust, $M_{\rm ISM}$), the mass in heavy elements (gas-phase metals and dust, $M_{\rm Z}$) and 
the dust mass ($M_{\rm d}$) in the diffuse (diff) and molecular cloud (MC) phase is described by the following system of equations:

\begin{equation}
\dot{M}_{\rm ISM}^{\rm mc}(t)  =  \dot{M}_{\rm cond}(t) - SFR(t) 
\label{eq:1}
\end{equation}
\begin{equation}
\dot{M}_{\rm ISM}^{\rm diff}(t)   = -\dot{M}_{\rm cond}(t)+\dot{R}(t) +\dot{M}_{\rm inf}(t) -\dot{M}_{\rm ej}(t)
\end{equation}
\begin{equation}
\dot{M}^{\rm mc}_{\rm Z} =  \dot{M}^{\rm Z}_{\rm cond}(t) - SFR(t)\,Z_{\rm mc} 
\end{equation}
\begin{equation}
\dot{M}^{\rm diff}_{\rm Z}  =  -\dot{M}^{\rm Z}_{\rm cond}(t) +\dot{Y}_{\rm Z}(t)
\end{equation}
\begin{equation}
\dot{M}^{\rm mc}_{\rm d}  =  \dot{M}^\mathcal{D}_{\rm cond}(t) - SFR(t)\,\mathcal{D}_{\rm mc} + \frac{M^{\rm mc}_{\rm d}(t)}{\tau_{\rm acc}} 
\end{equation}
\begin{equation}
\dot{M}^{\rm diff}_{\rm d}  =  -\dot{M}^{\mathcal{D}}_{\rm cond}(t) +\dot{Y}_{\rm d}(t) - \frac{M^{\rm diff}_{\rm d}(t)}{\tau_{\rm d}}, 
\label{eq:6}
\end{equation}
\noindent
where the time-dependent star formation rate ($SFR$) and the infall and outflow rates ($\dot{M}_{\rm inf}$ and $\dot{M}_{\rm ej}$)
are computed from the simulation outputs, $Z = M_{\rm Z}/M_{\rm ISM}$ and $\mathcal{D} = M_{\rm d}/M_{\rm ISM}$ are the ISM mass fractions in heavy elements and
dust, $\tau_{\rm d}$ and $\tau_{\rm acc}$ are the dust destruction and accretion timescales and will be defined below.
The terms  $\dot{M}_{\rm cond}, \dot{M}^{\rm Z}_{\rm cond}$, and $\dot{M}^\mathcal{D}_{\rm cond}$ describe 
the ISM, heavy elements and dust mass exchange between the diffuse phase and the molecular phase. Since these terms 
can not be directly inferred from
the simulation, we must resort to indirect constraints. By assuming that the SFR can be represented by
the Kennicutt-Schmidt relation, the mass of molecular gas can be estimated as,
\begin{equation}
M_{\rm ISM}^{\rm mc} = SFR(t) \frac{\tau_{\rm ff}}{\epsilon_{\ast}},
\end{equation}
\noindent
where $\epsilon_{\ast} = 0.01$ is the star formation efficiency \citep{Krumholz2012} and
\begin{equation}
\tau_{\rm ff} = \sqrt{\frac{3 \pi}{64\, G\, m_{\rm H} \, n_{\rm mol}}} 
\end{equation}
\noindent
is the free-fall timescale at the mean density of molecular clouds $\rho_{\rm mol} \sim 2 \, m_{\rm H} \, n_{\rm mol}$ \citep{Schneider2016}, and
we assume $n_{\rm mol} = 10^3$~cm$^{-3}$ to be consistent with the value adopted for the grain growth timescale (see below). 
We use the above condition to compute $\dot{M}_{\rm cond}$. 
The mass exchange of heavy elements and dust 
depends on the degree of enrichment of each phase. Hence, we compute $\dot{M}^{\rm Z}_{\rm cond}$ by requiring the heavy
elements abundance in the molecular clouds, $Z_{\rm mc}$, to be equal to the metallicity of newly formed stars predicted by
the simulation, $Z_\ast$, and we assume that $\dot{M}^{\mathcal{D}}_{\rm cond} = \mathcal{D}/Z \, \dot{M}^{\rm Z}_{\rm cond}$.
Finally,  $\dot{R}$, $\dot{Y}_{\rm Z}$ and $\dot{Y}_{\rm d}$ are, respectively, the return mass fraction, the yields of heavy elements and the dust yields 
produced by stellar sources and depend on the SFR, 
the IMF, and on the adopted metal and dust stellar yields. Following \citet{Valiante2009} we compute them as:
\begin{equation}
\dot{R}(t)=\int^{m_{\rm up}}_{m(t)} (m-w_m(m, Z))\, \Phi(m) \, {\rm SFR}(t-\tau_{\rm m}) \,dm,
\end{equation}
\noindent
\begin{equation}
\dot{Y}_{\rm Z}(t)=\int^{m_{\rm up}}_{m(t)} m_{\rm Z}(m, Z)\, \Phi(m) \, {\rm SFR}(t-\tau_{\rm m}) \,dm,
\end{equation}
\noindent
and
\begin{equation}
\dot{Y}_{\rm d}(t)=\int^{m_{\rm up}}_{m(t)} m_{\rm d}(m, Z)\, \Phi(m) \, {\rm SFR}(t-\tau_{\rm m}) \,dm,
\end{equation}
\noindent
where   $\tau_{\rm m}$ is the lifetime of a star with mass 
$m$, $\Phi(m)$ is the IMF, $w_m, m_{\rm Z},$ and  $m_{\rm d}$ are, respectively, the mass of the stellar remnant, of heavy elements and dust produced by a star 
with mass $m$ and metallicity $Z$, and the integral is computed from the upper mass limit of the IMF down to the mass that has a lifetime $\tau_{\rm m} = t$. For stars with mass $m < 8 M_\odot$ we adopt dust yields from \citet{Zhukovska2008}, while for stars in the mass range $12$ $\rm M_\odot$ to $40$ $\rm M_\odot$ dust yields are taken from \citet{Bianchi2007}, including the effect of dust destruction by the SN reverse shock \citep{Bocchio2016}. For stars in the intermediate mass range $8 \, M_\odot < m < 12 \, M_\odot$ the dust yields are computed interpolating from the values corresponding to the most massive AGB progenitor and the least massive SN progenitor. Finally, stars with $m > 40 \, M_\odot$ are assumed to collapse to a black hole, 
without enriching the surrounding ISM. The other terms in the right-hand side of equations (\ref{eq:1})-(\ref{eq:6}) account for the effects of astration and dust reprocessing in the ISM.
The dust destruction timescale is modelled as in \citet{Valiante2011} and \citet{deBennassuti2014}, 
\begin{equation}
\tau_{\rm d} = \frac{M_{\rm ISM}^{\rm diff}}{R_{\rm SN}^\prime \epsilon_{\rm d} M_{\rm s}(v_{\rm s})}, 
\end{equation}
\noindent
where $M_{\rm s}(v_{\rm s}) = 6800 M_{\odot} \langle E_{51} \rangle /(v_{\rm s}/100 {\rm km/s})^2$
is the mass shocked to a velocity of at least $v_{\rm s}$ by a SN in the Sedov-Taylor phase,
$R_{\rm SN}^\prime$ is the {\it effective} SN rate, since not all SNe are equally efficient at destroying dust \citep{McKee1989},
$\epsilon_{\rm d}$ is the dust destruction efficiency, and $\langle E_{51} \rangle$ is the average SN energy in units of $10^{51}$~erg. In what follows, we assume 
$R_{\rm SN}^\prime = 0.15 R_{\rm SN}$, where $R_{\rm SN}$ is the core-collapse SN rate, $\langle E_{51} \rangle = 1.2$, $v_{\rm s} = 200$~km/s,
and $\epsilon_{\rm d} = 0.48$ \citep{Nozawa2006}.

The grain growth timescale is parametrized as in \citet{mancini2015},
\begin{eqnarray}
\tau_{\rm acc}=	2 \, {\rm Myr} \times \left(\frac{n_{\rm mol}}{1000 \, {\rm cm}^{-3}}\right)^{-1} 
					\left(\frac{T_{\rm mol}}{50 \, {\rm K}}\right)^{-1/2}
					\left(\frac{Z}{Z_\odot}\right)^{-1}\\ \nonumber
					=\tau_{\rm acc,0} \left(\frac{Z}{Z_\odot}\right)^{-1}					 
 \label{tauaccr}
 \end{eqnarray}
 \noindent
where we have assumed that the grains have a typical size of $\sim 0.1 \mu {\rm m}$ \citep{Hirashita2014a} and $n_{\rm mol}$, $T_{\rm mol}$ are the number density and temperature of the cold molecular gas phase where grain growth is more efficient. Since these values refer to structures that are not resolved  by the simulation, we assume $n_{\rm mol} = 10^3$~cm$^{-3}$, $T_{\rm mol} = 50$~K which lead to a constant value of $\tau_{\rm acc,0} = 2 \,{\rm Myr}$. These conditions have been shown to reproduce the observed dust-to-gas ratio of local galaxies over a 
wide range of metallicities \citep{deBennassuti2014}. In addition,  \citet{mancini2015} show that with these parameters, the model predictions are consistent with the upper limits on the dust mass inferred from deep ALMA and PdB observations of galaxies at $z > 6$ (see the observations shown in Fig.~\ref{fig:dustm}). As
pointed out by \citet{mancini2015}, the observed dust mass in the galaxy A1689-zD1 at $z=7.5$ can only be explained using a value of $\tau_{\rm acc,0} = 0.2$~Myr (see also Section 3.1).

\begin{figure}
\includegraphics[width=.9\columnwidth]{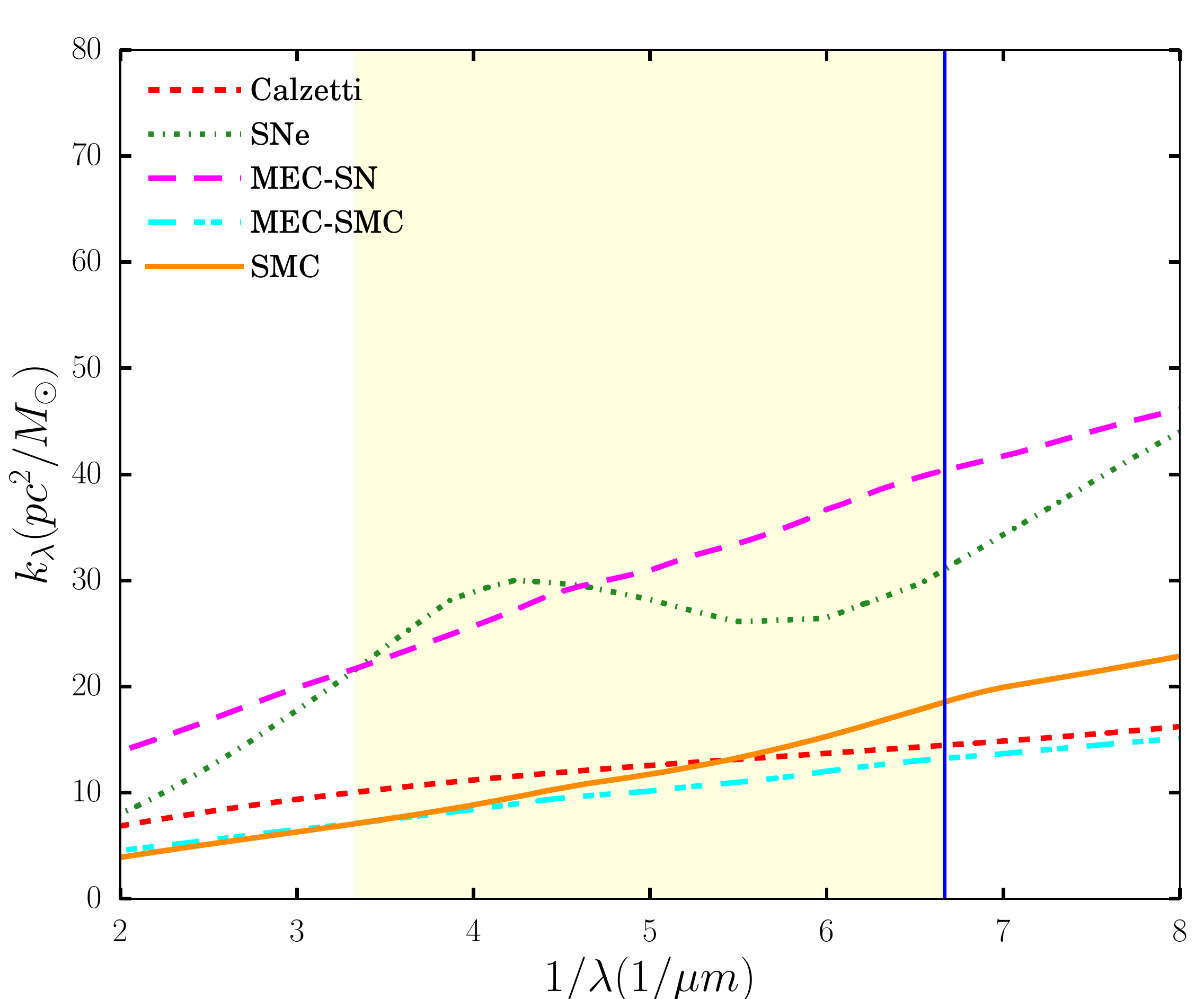}
\caption{Extinction coefficient per unit dust mass as a function of $1/\lambda$. 
The lines represent different grain models:
SMC (orange solid), Calzetti (red dashed, normalized at $\lambda = 3000 \, \AA$ to the Milky Way extinction curve), SN (dark green dot-dashed) and MEC
normalized at $\lambda = 3000 \, \AA$ to the SN extinction curve (magenta long-dashed) and to the SMC curve (cyan dot-dashed).
The vertical solid blue line corresponds to the wavelength  $\lambda = 1500 \, \MAA$ at which we compute the galaxy 
rest-frame luminosity, and the shaded region is the rest-frame UV wavelength range 
$1500 \, \MAA \leq \lambda \leq 3000 \, \MAA$ used to compute the $\beta$ slopes (see text).}
\label{fig:extinctionCurve}
\end{figure}

\begin{figure*}
\includegraphics[width=\textwidth]{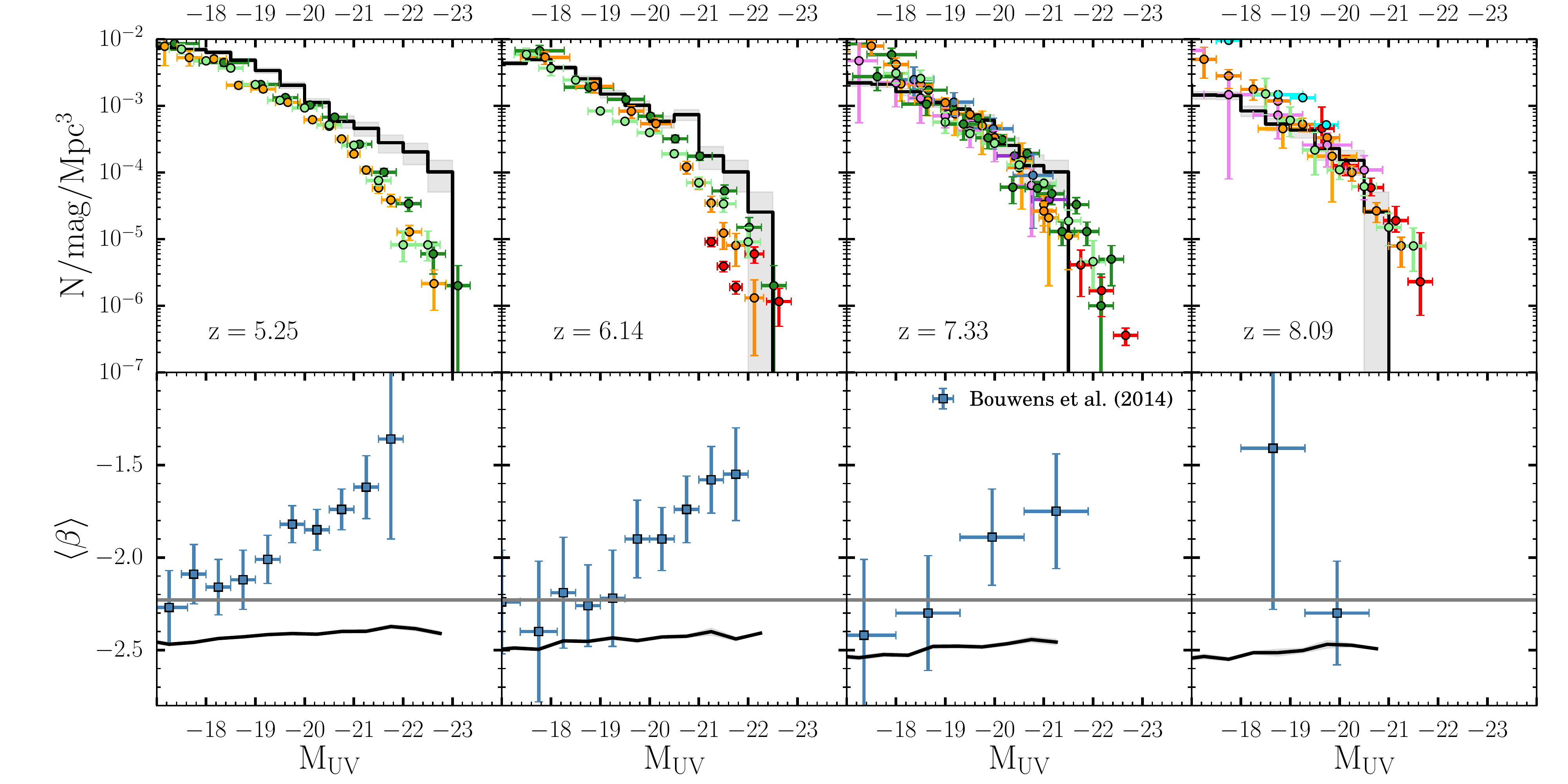}
\caption{ {\it Top Panel:} comparison between the {\it intrinsic} UV LF of the simulated galaxies (black lines) and the 
observations at redshift $z\sim 5, 6, 7 $ and 8 (from left to right). The data are taken from
\citet{mclure2009} (orange), \citet{oesch2010} (dark blue), \citet{Bouwens2015} (dark green),
\citet{Finkelstein2015} (light green), \citet{castellano2010} (dark violet), \citet{mclure2010} (light orange), \citet{McLure2013} (dark orange), \citet{Atek2015} (cyan), \citet{Laporte2015} (light violet), and  \citet{Bowler2014, Bowler2015} (red). The shaded regions indicate Poissonian errors.
{\it Bottom Panel}: mean spectral index $\left\langle \beta \right\rangle$ as function of the {\it intrinsic} magnitude $M_{\rm UV}$ at the same redshifts. The model prediction for the intrinsic colours are shown as solid black lines with shaded regions indicating the standard errors on the mean.
Blue squares indicate the observations by \citet{Bouwens2014}. The horizontal solid grey line shows the value $\beta = -2.23$ that is adopted in the \citet{meurer1999} relation (see text). 
A coloured version of this figure is available online.}
\label{fig:combint}
\end{figure*}

\subsection{Modeling the extinction}
\label{met:extinction}

The radiation flux escaping a galaxy can be derived from its intrinsic emission, given by eq.~(\ref{eq:int_sed}), and accounting for the wavelength dependent
extinction of the galactic ISM. 
In our computational scheme, stellar populations are represented by stellar particles which can experience different absorptions/obscurations 
depending on the columns of dust in their surroundings. For this reason, we compute the flux emerging from the $j$-th galaxy by applying a different extinction to 
each stellar particles $i$, and then summing up their contribution to obtain the total escaping flux:
\begin{equation}
\bar{F}^j_\lambda=\sum_i {\cal K}^i(\tau_\lambda) \,  f^i_\lambda \, M_{\ast,i},
\label{eq:ftot}
\end{equation}
\noindent
where ${\cal K}^i(\tau_\lambda)$ is the extinction 
factor per stellar particle, as a function of the optical depth $\tau_\lambda$ at a specific wavelength $\lambda$.
Note that the modeling of ${\cal K}^i(\tau_\lambda)$ is a complicated task because of the multi-phase nature of the ISM: both the medium surrounding each stellar particle and the diffuse ISM contribute to the extinction. Hence,
${\cal K}^i(\tau_\lambda)$ depends on the dust content and on its spatial distribution relative to the stars.  During their lifetime, stars evolve, changing their intrinsic SED, and interact with their environment through mechanical, chemical and radiative feedback effects. For these reasons, the values of ${\cal K}^i(\tau_\lambda)$ experienced by stellar populations change with time.

In our reference model, we assume that all stars form in molecular clouds, from which they escape in a typical timescale $t_{\rm esc}$\footnote{This value can be also interpreted as the molecular cloud dissipation timescale.},  moving into the diffuse phase. Hence, if the age of the stellar population is $t_{\star}<t_{\rm esc}$, the emitted radiation is extinguished by the additional column of dust of the parent molecular cloud, namely:

\begin{eqnarray}
\tau_{\lambda} =  
\tau^{\rm mc}_{\lambda} +\tau^{\rm diff}_{\lambda} & \textit{if  }\,\,\,
t_{\rm \star}<t_{\rm esc} \nonumber \\
\tau_{\lambda} =  
\tau^{\rm diff}_{\lambda}& \textit{if  } \,\,\,
t_{\rm \star}\ge t_{\rm esc}.
\label{eq:tau}
\end{eqnarray}

A similar model was originally
proposed by \citet{Charlot2000} as an idealised description of the ISM to compute the effects of dust on the integrated spectral properties of galaxies, and it
has been applied by \citet{ForeroRomero2010} to describe the clumpy structure of the ISM in high-$z$ galaxies.
The observed flux is finally computed solving the radiative transfer equation in both phases, by assuming a homogeneous, one-dimensional and isotropic gas/dust distribution, i.e. :

\begin{eqnarray}
	\rm	{\cal K}^i(\tau_\lambda) =  e^{-(
	\tau^{mc}_\lambda +\tau^{diff}_\lambda)}& \textit{if  } \,\,\,
	t_{\rm \star}<t_{\rm esc} \nonumber \\
	\rm	{\cal K}^i(\tau_\lambda) = e^{-\tau^{diff}_\lambda}& \textit{if  } \,\,\,
	t_{\rm \star} \geq t_{\rm esc}.
\label{eq:obsflux}
\end{eqnarray}
\noindent
More details on the calculation of $\tau_\lambda$ for both phases can be found in the next section.

\subsection{Dust optical depth}
The optical depth at a fixed $\lambda$ depends both on the type of absorbers present in the medium and on their column density. 
Observations of high-$z$ galaxies probe the restframe UV range, where the radiation is mostly 
extinguished by dust and the optical depth can be computed as,
\begin{equation}
\tau_{\lambda}=\Sigma_{\rm d} \, k_\lambda,
\end{equation}
\noindent
where $\Sigma_{\rm d}$ is the dust column density and $k_\lambda$ is the extinction coefficient per unit dust mass. 
We assume that molecular clouds can be approximated as spheres of constant mass $M_{\rm cloud}$ and volume 
density $n_{\rm mol}$, and that their surface density can be expressed as,
\begin{equation}
\Sigma^{\rm mc}_{\rm ISM} = 9.9 \times 10^2 \, \frac{M_{\odot}}{\rm pc^2} \,\, \left(\frac{M_{\rm cloud}}{10^{6.5}\; M_{\odot}}\right)^{1/3}
\left(\frac{n_{\rm mol}}{1000 \; {\rm cm}^{-3}}\right)^{2/3},
\end{equation}
\noindent
where we adopt $n_{\rm mol} = 1000 \, \rm cm^{-3}$ (see Section \ref{met:chem}) and a cloud mass of $M_{\rm cloud} = 10^{6.5}\, M_{\odot}$,
which corresponds to the typical mass of the largest giant molecular clouds observed in the Milky Way \citep{Murray2011}.
Following \citet{Hutter2014}, we compute the diffuse gas column density as,
\begin{equation}
\Sigma^{\rm diff}_{\rm ISM}  =  \frac{M^{\rm diff}_{\rm ISM} }{\pi r^2_{\rm d}}  
\end{equation}
\noindent
and the radius of the gas distribution as $r_{\rm d} = 4.5 \,  \lambda \,  r_{\rm vir}$ \citep{Ferrara2000},
where $r_{\rm vir}$ is the dark matter halo virial radius and $\lambda = 0.04$ is the mean value of the dark matter halo spin distribution. 
Finally, under the assumption that dust is uniformly mixed with the gas, the dust surface densities in the two phases can be derived as
$\Sigma^{\rm mc}_{\rm d} = \mathcal{D}_{\rm mc} \, \Sigma^{\rm mc}_{\rm ISM}$ and $\Sigma^{\rm diff}_{\rm d} = {\cal D}_{\rm diff} \, \Sigma^{\rm diff}_{\rm ISM}$.

The dust extinction coefficient, $k_\lambda$, depends on the grain size distribution and on the optical properties of the grain species. Unfortunately, we still lack
a model that is able to self-consistently predict the evolution of the dust mass and extinction properties. In the local Universe,
the average dust extinction properties of the Milky Way, the Large and Small Magellanic Clouds are different, probably
as a result of their different star formation and chemical evolution histories \citep{Cardelli1989, Pei1992, Weingartner2001}.
At high redshifts, the SED of star forming galaxies is generally modeled with the \citet{Calzetti1994} attenuation law, although 
a steeper extinction curve, such as the SMC, often provides a better description \citep{Reddy2010}. Using a sample of quasars at $3.9 \leq z \leq 6.4$, \citet{Gallerani2010} inferred
a mean extinction curve that is flatter than the SMC curve which is generally applied to quasar at $z < 4$. They discussed
the possibility that this difference may indicate either a different dust production mechanism
at high redshift, or a different mechanism for processing dust into the ISM and suggested that the same transitions may also apply to
normal, star-forming galaxies at $z > 4$. Indeed, at $z \sim 6$ evidence
of an extinction law very similar to the one predicted by theoretical models for dust formed in SN ejecta has been found in
the spectra of the reddened quasar SDSSJ1048+46 at $z = 6.2$ \citep{Maiolino2004},
the GRB050904 afterglow at $z = 6.3$ \citep{Stratta2007} and the GRB071025 at $z \sim 5$ (\citealt{Perley2010}, see however 
\citealt{Zafar2010} for a different conclusion). 

Since we do not know how the dust extinction properties change with redshift, here we consider four 
different extinction curves, that we show in Fig.~\ref{fig:extinctionCurve}: the SMC extinction curve 
\citep{Weingartner2001,Pei1992}, the Calzetti model \citep{Calzetti2000}, the
extinction curve derived for grains formed in SN ejecta \citep{Bianchi2007}, and the mean extinction curve (MEC)
inferred by \citep{Gallerani2010}\footnote{The MEC and Calzetti attenuation models have been normalized 
at $\lambda = 3000 \, \AA$ to the values predicted by the SN and SMC and by the Milky Way extinction curves, respectively.}
The vertical solid line indicates the value $\lambda = 1500 \,\MAA$ at which we compute the galaxy restframe UV luminosity,
and the shaded region identifies the wavelength range where we compute the $\beta$ slopes. In this range, the Calzetti, SMC and MEC models
show a smooth increase with $\lambda^{-1}$, although with a different slope. Conversely, the SN extinction curve shows
a spectral bump due to amorphous carbon grains \citep{Bianchi2007}. 
Overall, we expect the MEC curve normalized to the SN extinction coefficient at $\lambda = 3000 \, \AA$ to have the largest effect on the restframe UV colours.
In fact, the extinction coefficient per unit dust mass for this model at $1500 \, \AA$ is a factor $\approx 2.7$ larger than the one predicted by the Calzetti extinction
curve.

\begin{figure}
\hspace{-0.1\columnwidth}
\includegraphics[width=1.1\columnwidth]{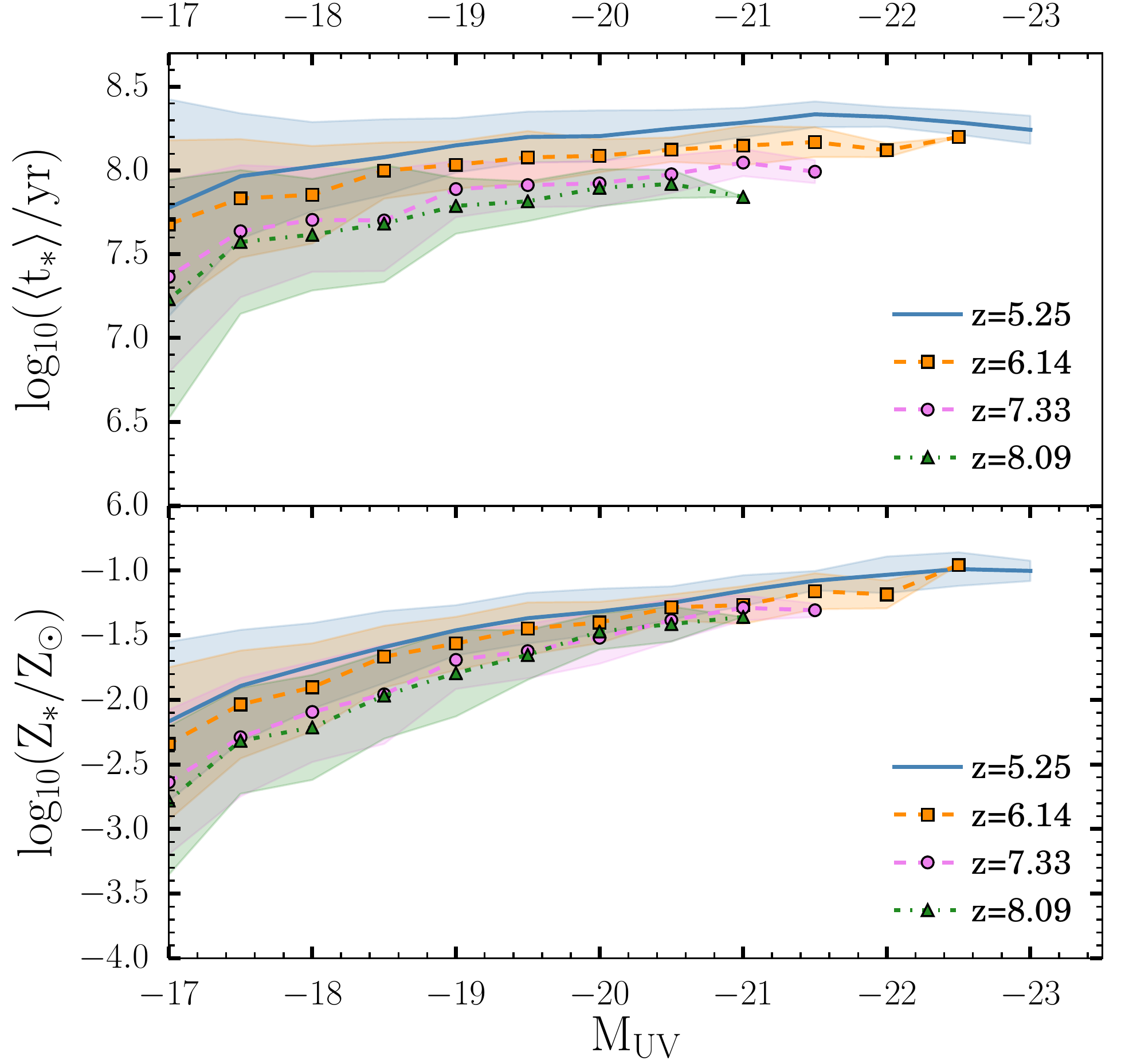}
\caption{
Mass-averaged stellar age (top panel) and metallicity (bottom panel) of the simulated galaxies as a function of their intrinsic UV 
magnitude ($\lambda=1500 \MAA$). The different lines indicate the mean values at redshift $\sim 5$ (blue solid), 6 (orange dashed with squares), 7 (magenta dashed 
with dots) and 8 (green dotted with triangles). The shaded regions indicate the 1-$\sigma$ deviation.
A coloured version of the figure is available online.}
\label{fig:stage}
\end{figure}

\begin{figure*}
\includegraphics[width=\textwidth]{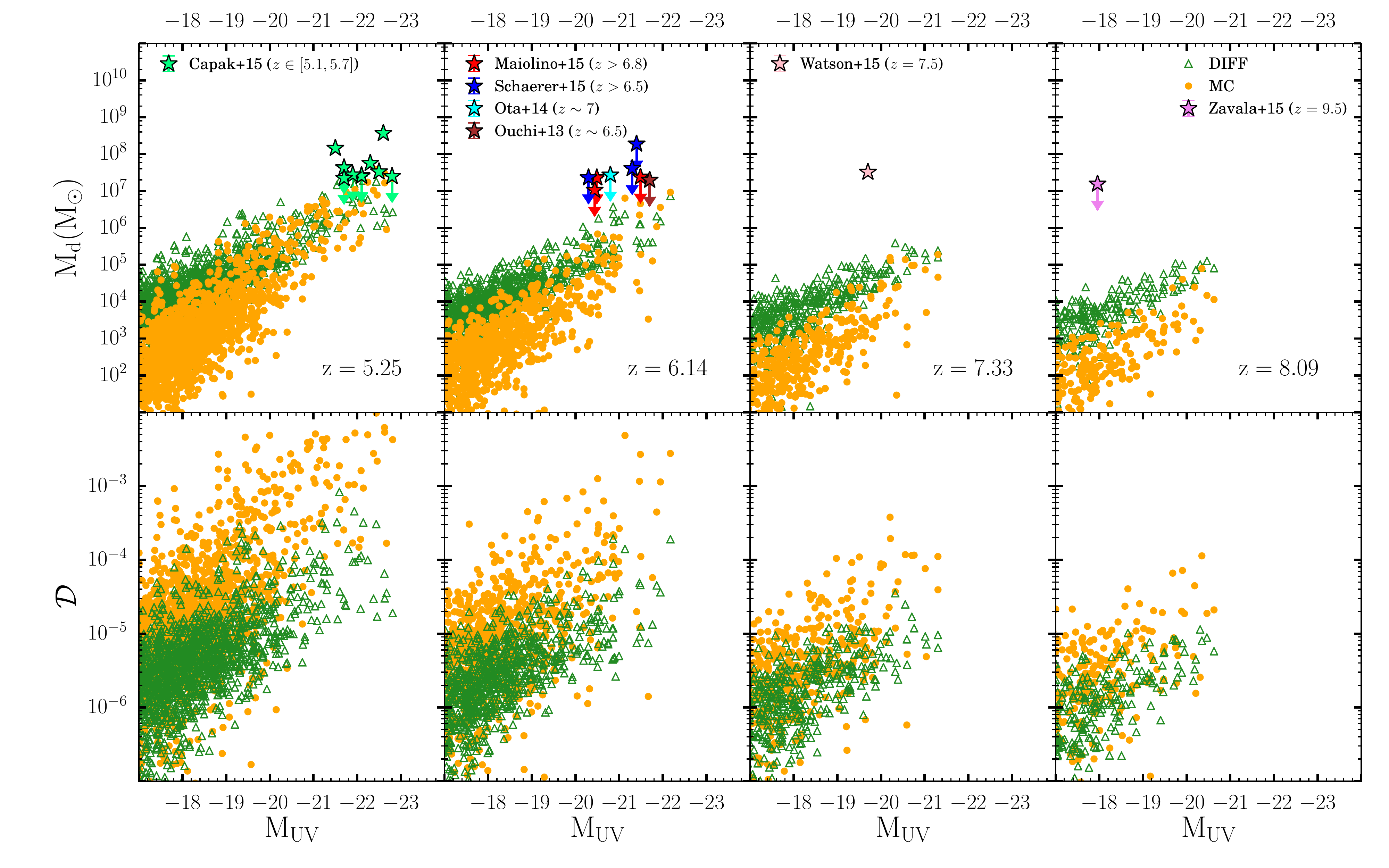}
\caption{The dust mass ({\it top panels}) and dust-to-gas mass ratio ({\it bottom panels}) as a function of the intrinsic UV magnitude of the simulated galaxies
at $z \sim 5, 6, 7,$ and 8 (from left to right). In each panel, we show the dust mass and dust-to-gas mass ratio in molecular clouds (orange dots) and in the diffuse 
phase (green triangles). In the upper panel we also show data and upper limits from \citet{Kanekar2013}, \citet{Ouchi2013}, \citet{Ota2014}, \citet{Capak2015}, \citet{Schaerer2015}, \citet{Maiolino2015}, \citet{Watson2015},
and \citet{Zavala2015}, see text.}
\label{fig:dustm}
\end{figure*}

\begin{figure*}
\includegraphics[width=\textwidth]{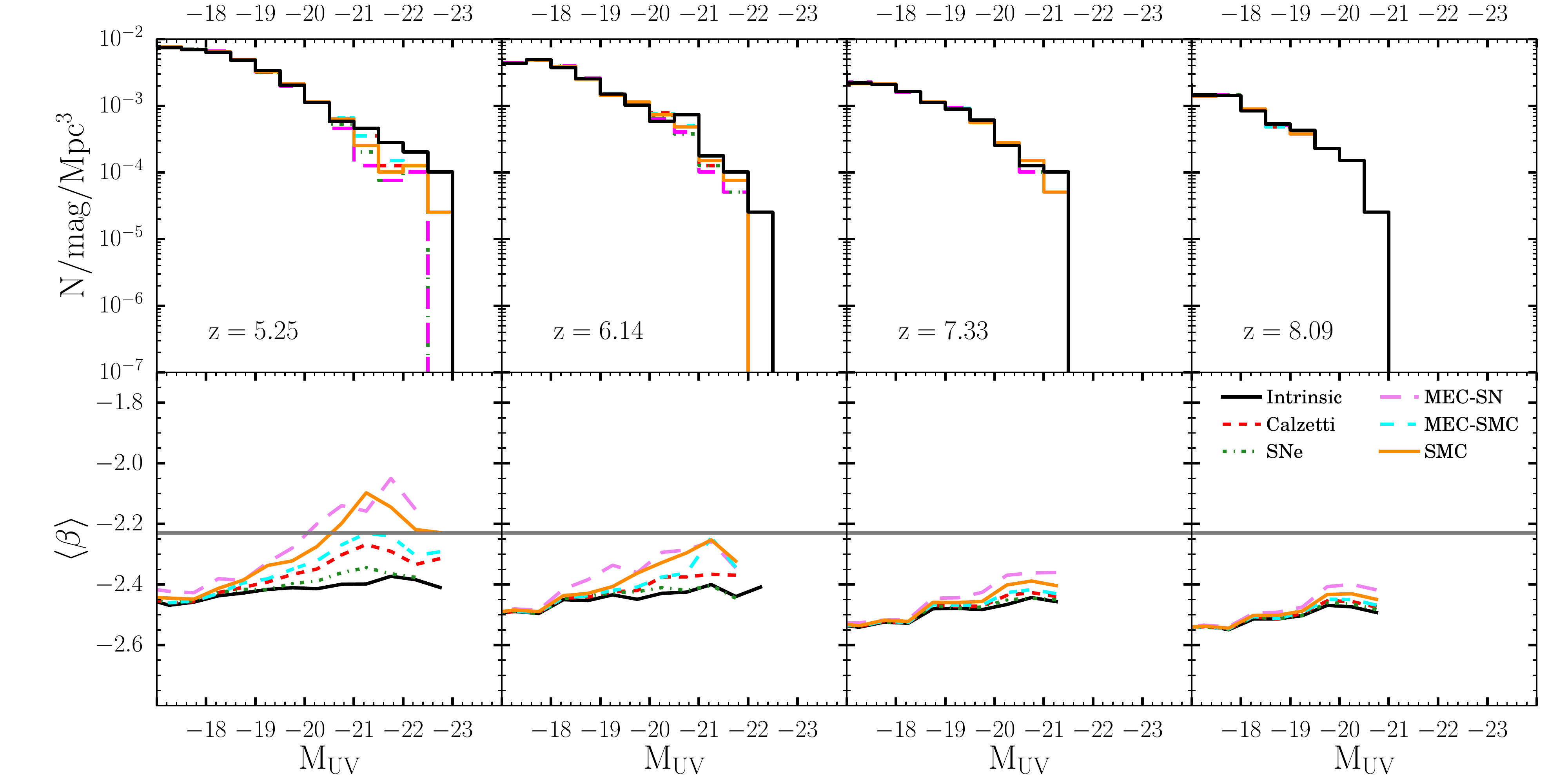}
\caption{Predicted UV LF ({\it top panels}) and $\beta$ slopes ({\it bottom panels}) 
at $z = 5, 6, 7,$ and 8 (from left to right) assuming $t_{\rm esc} = 0$ and different dust extinction curves,
colour-coded as in Fig.~\ref{fig:extinctionCurve}.
In each panel, we also report the {\it intrinsic} LFs and $\beta$ shown in Fig.~\ref{fig:combint} (black solid lines). The
horizontal grey lines in the bottom panels show the value $\beta = -2.23$ adopted in the \citet{meurer1999} relation.
A coloured version of this figure is available online.}
\label{fig:comboLfBeta}
\end{figure*}

\begin{figure*}
\includegraphics[width=\textwidth]{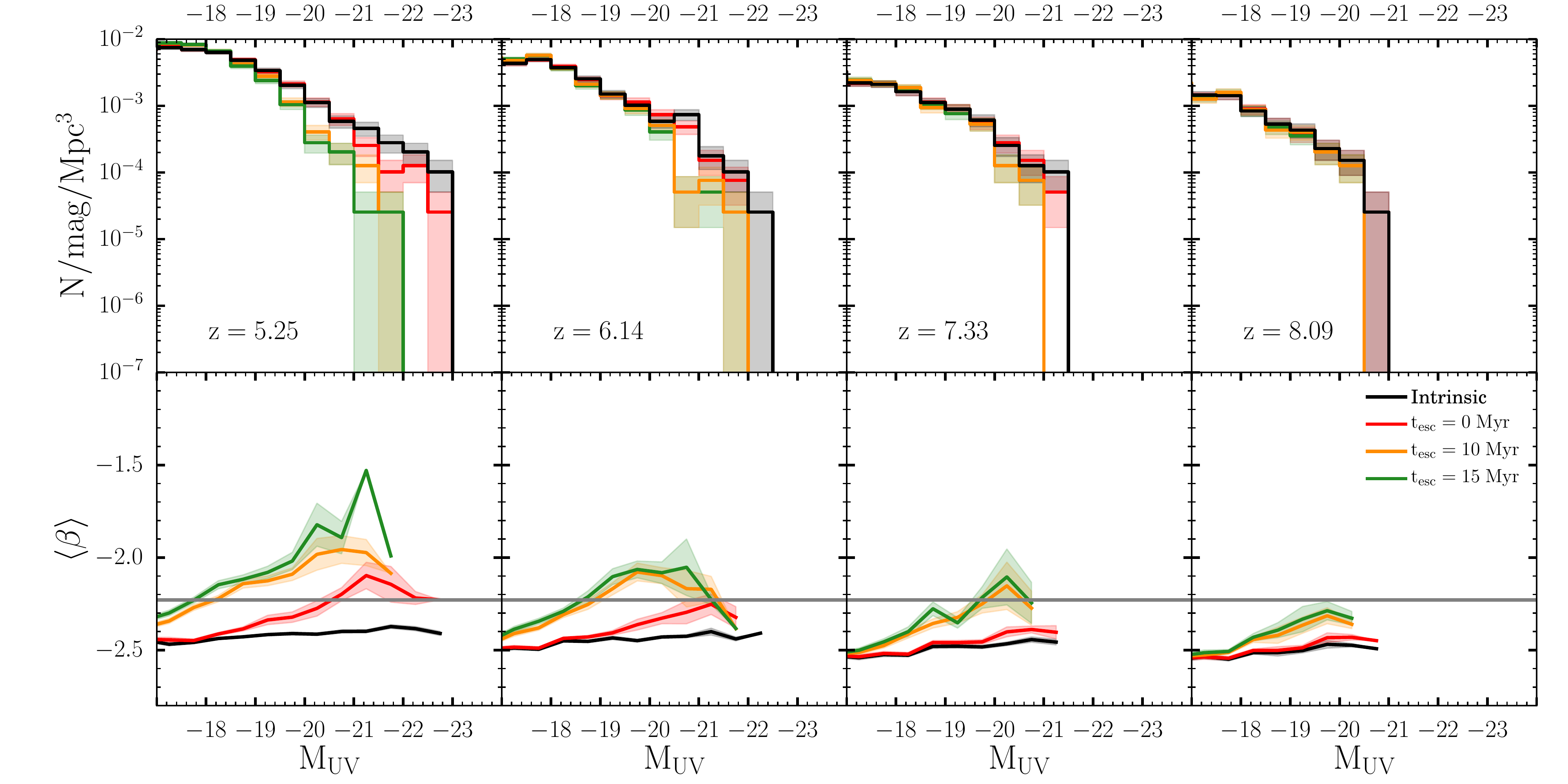}
\caption{Same as in Fig.\ref{fig:comboLfBeta}, but assuming an SMC extinction curve and varying the parameter $t_{\rm esc}$ from
0 to 15 Myr. The shaded regions represent Poissonian errors (top panels) and standard errors on the mean (bottom panels).}
\label{fig:lftim}
\end{figure*}

\section{Results}
\label{sec:results}

In this section, we first present the physical properties of the simulated galaxies at $5 \le z \le 8$, and we compute their {\it intrinsic} UV
luminosities and $\beta$ slopes. We then explore the effects of dust extinction as predicted by different extinction models.

\subsection{Physical properties of early galaxies}
\label{res:physprop}

For each simulated galaxy, we first compute the absolute UV magnitude at $1500 \,\AA$, $M_{\rm UV}$, and the $\beta$ slope in the
wavelength range $[1500 - 3000]\,\AA$, from the SED presented in Section \ref{met:spectra}.
Hence, we assume that the UV emission produced by the stellar populations does not suffer any extinction from interstellar dust. At each
redshift, we distribute galaxies in different magnitude bins and compute the resulting UV LF and the average colour, $\langle \beta \rangle$.

Fig.~\ref{fig:combint} 
shows a comparison between the model predictions and a collection of observational data taken from the literature (see the caption for details). 
 At $z \sim 7 - 8$, the {\it intrinsic} LFs underpredict the number of galaxies at the bright-end (with $M_{\rm UV} \leq -21$) and at the faint-end 
(with $M_{\rm UV} \geq -18$). This is an effect of the limited volume size and mass resolution of the simulation (see also \citealt{salvaterra2013}).
 At lower $z$, the effect of numerical resolution is smaller, but we do not find any galaxy with {\it intrinsic}
$M_{\rm UV} \le -23$ in the simulated volume. In a future study, we plan to apply this analysis to a new simulation 
with a larger box size and a comparable mass resolution. In fact, our main interest is to increase the
statistics at the high-mass end, where we expect the effects of dust extinction to be more prominent. 
Within these limitations, at $z <7$ the number of galaxies at the bright-end 
 is larger than observed, as already discussed by \citet{salvaterra2013,dayal2013, Finkelstein2015} and \citet{Khakhaleva2016}. We find that, {\it while the observed LF at $z \sim 8$ is consistent with
 negligible dust extinction, at $z \sim 5$ observations seem to require significant dust extinction at all luminosities brighter than $\rm M_{UV} \sim - 18$.}
  
At all redshifts, the predicted {\it intrinsic} $\langle \beta \rangle$ slopes are much bluer than observed, particularly at the bright end (bottom panels). The simulated galaxies
show similar colours at all luminosities, with $\left\langle \beta \right\rangle \sim -2.5$, and only a very modest increase with cosmic time. 
These colours are bluer than the $\beta = -2.23$ adopted in the \citet{meurer1999} relation (shown by the horizontal grey line), as already noticed by \citet{Wilkins2012b}.

The above trends can be easily understood by looking at the mean physical properties of the stellar populations in the simulated galaxy samples.
Fig.~\ref{fig:stage} shows the mass-averaged stellar age (upper panel) and metallicity (bottom panel) of the simulated galaxies at $5 \leq z \leq 8$
as a function of their {\it intrinsic} UV magnitudes. We find that galaxies with a given luminosity tend to be slightly younger and less metal-enriched 
at higher redshift ($80~{\rm Myr} \leq \langle t_* \rangle \leq 160 \,{\rm Myr}$ and  $0.03 \, Z_\odot \leq \langle Z_* \rangle \leq 0.06 \, Z_\odot$ 
for galaxies with $M_{\rm UV} = -20$ and $5\leq z \leq 8$), 
and that the average stellar age and metallicity increases with the {\it intrinsic} UV luminosity,
showing a larger dispersion of values for fainter galaxies with $M_{\rm UV} \geq -18$. However, their overall properties do not show
 a significant evolution with UV magnitude and redshift, 
consistent with their relatively constant {\it intrinsic} UV colours. 

Over the same UV luminosity and redshift
range, the dust mass in their ISM varies significantly. This is shown in Fig.~\ref{fig:dustm}, where we plot the dust
mass, $M_{\rm d}$, derived as explained in Section \ref{met:chem}, and the dust-to-gas mass ratio, ${\cal D}$, 
as a function of the {\it intrinsic} UV magnitudes. The dust mass increases with UV luminosity and - for a given luminosity - galaxies 
become more dust-enriched with cosmic time.  As already discussed in \citet{mancini2015}, the dust mass increases with stellar mass,
hence with the {\it intrinsic} UV luminosity. In low-mass galaxies, the dust mass has mostly a stellar origin (SNe and AGB stars). In
massive and chemically evolved galaxies, grain growth becomes progressively more efficient, providing the dominant contribution to the total dust mass.
Hence, the dust mass in the molecular phase increases with galaxy luminosity and becomes as large as the dust mass in the diffuse phase for galaxies
with {\it intrinsic} $\rm M_{UV} < - 20$. 

In Fig.~\ref{fig:dustm} we also show the data and upper limits on the dust mass inferred
from observations at $z \sim 5.1 - 5.7$ \citep{Capak2015}, $z \sim 6.5 - 7$ \citep{Kanekar2013, Ouchi2013, Ota2014, Schaerer2015, Maiolino2015}, 
$z \sim 7.5$ \citep{Watson2015} and $z \sim 9.6$ (\citealt{Zavala2015}, that we arbitrarily report in the $z \sim 8$ panel). Following \citet{mancini2015},
we have estimated the dust mass from the observed mm flux assuming optically thin emission, a dust emissivity $k_{\nu_{\rm res}} = k_0 (\lambda_0/\lambda_{\rm res})^\beta$
(with $k_0 =   0.77\, {\rm cm}^2/{\rm gr}$, $\lambda_0 = 850 \,\mu$m and $\beta = 1.5$, \citealt{Ota2014}), and a dust temperature of 35~K. 
The resuting dust masses (shown as  starred data points) have been reported only for indicative purposes, as a more meaningful comparison between model predictions and observations is given in Section \ref{sec:comparison}. While the predicted dust masses
are consistent with observations at $z \sim 5.1 - 5.7$ and with the upper limits inferred at $z > 6$, the data reported by \citet{Watson2015}, and recently
confirmed with deeper observations by \citet{Knudsen2016}, on the $z = 7.5$ galaxy A1689-zD1, requires more efficient grain growth, as if the galaxy were characterized
by a denser ISM \citep{mancini2015, Michalowski2015}.

Due to their different gas content, the average dust-to-gas mass ratio is smaller in the diffuse phase than in
molecular clouds and, in both phases, ${\cal D}$ grows with UV luminosity. Hence, we expect the most massive galaxies, with the largest {\it intrinsic} UV luminosity, to 
experience a larger degree of dust extinction. Yet, ${\cal D}$ shows a large dispersion, particularly in the molecular phase, and galaxies
with the same {\it intrinsic} UV luminosity can be characterized by values of ${\cal D}$ which differs by 2 - 3 orders of magnitudes, particularly at lower $z$. 
At $z \lesssim 6$, galaxies with $M_{\rm UV} \leq -21$ have dust masses  which range between $10^6 \, M_\odot$
and $\approx 5 \times 10^7 \, M_\odot$ and dust-to-gas mass ratios in the molecular phase that can reach values of 
${\cal D} \geq 10^{-3}$. Conversely, at $z \gtrsim 7$, most of the simulated
galaxies have $M_{\rm d}  <10^6 M_\odot$ and ${\cal D} < 10^{-4}$. On the basis of these results we expect dust extinction to be more
relevant for bright galaxies, particularly at $z \lesssim 6$, where the deviations between the {\it intrinsic} UV LF and the data, shown in Figure \ref{fig:combint}, are more significant.

\subsection{The effects of dust extinction on the UV luminosities and colours}
\label{res:physdep}

We first consider the simplest model of dust extinction from the diffuse phase only, computing the optical depth
using Eqs.~(\ref{eq:tau})-(\ref{eq:obsflux}) with $t_{\rm esc} = 0$. In Fig.~\ref{fig:comboLfBeta}, we show the 
UV LF and CMR assuming the SMC, the Calzetti, the SN and the 
MEC extinction curves. For reference, we also show the {\it intrinsic} UV LF and $\langle \beta \rangle$ colours discussed
in the previous section, assuming no dust extinction. It is clear that dust extinction decreases the number of 
galaxies at the bright end, particularly at $z \leq 6$. The strongest effect is achieved using the MEC-SN and SN 
curves, as these models predict the largest $k_\lambda$ at $\lambda = 1500 \, \MAA$, followed by the SMC, the 
Calzetti and the MEC-SMC curves (see Fig.~\ref{fig:extinctionCurve}). These different models have an even larger
effect on the CMR, as this is sensitive
to the shape of the extinction curve over the wavelength range $1500\, \MAA \leq  \lambda \le 3000 \,\MAA$.
In fact, while the SN, the Calzetti, and the MEC-SMC models introduce only a mild reddening in the predicted colours, 
the SMC and MEC-SN models increase the $\left\langle \beta \right\rangle$ creating a dependence on the UV magnitude,
with the brightest galaxies being redder than the fainter ones. 
Hence, this analysis shows that {\it estimating dust attenuation from the observed $\beta$ can lead to very different results depending on
the adopted extinction curve. A flat extinction curve in the UV can hide a significant mass of dust under
relatively blue colours}. Overall, we find that - due to the low ${\cal D}$ of the diffuse phase
(see Fig.~\ref{fig:dustm}), when $t_{\rm esc} = 0$ dust extinction introduces only a modest reddening to the UV colours
and it has a negligible effet on the LFs at $z \gtrsim 6$.  

We finally discuss the effects of dust extinction on young stellar populations that are still embedded in their parent molecular clouds,
assuming $t_{\rm esc} = 10$ and 15 Myr in Eqs.~(\ref{eq:tau})-(\ref{eq:obsflux}). The results are shown in Fig.~\ref{fig:lftim},
where we have adopted the SMC extinction curve. For reference, we also report in the same figure the {\it intrinsic} UV LF and
colours as well as the results discussed above, when $t_{\rm esc} =0$. Not surprisingly, {\it the longer the time 
young stellar populations spend in their natal molecular clouds, the largest is the effect of dust extinction, both
on the bright-end of the luminosity function and on the $\beta$ slopes} (see also \citealt{ForeroRomero2010}).  Due to the larger values of ${\cal D}$ in the
molecular phase, the predicted $\left\langle \beta \right\rangle$ for $M_{\rm UV} = -20$ galaxies at $z \sim 8$ increases from 
$\sim -2.5$ when $t_{\rm esc} = 0$ (essentially the {\it intrinsic} $\beta$ value) to $\sim -2.3$ when $t_{\rm esc} = 10 - 15$ Myr.
The increasing efficiency of grain growth causes $M_{\rm UV} = -20$ galaxies at $z \sim 5$ to have $\left\langle \beta \right\rangle = - 2.3$
when $t_{\rm esc} = 0$ and as large as $-2$ ($-1.8$) when $t_{\rm esc} = 10$ ($15$) Myr.

\section{Comparison with observations}
\label{sec:comparison}
In this section, we first compare the model predictions with the observed UV LFs and CMR. Then, we analyze the origin of the scatter
around the CMR and the stellar mass - UV luminosity relation at different $z$. Finally, we compute the IRX and dust attenuation factors
as a function of $\beta$. 

\subsection{UV luminosity function and Colour-Magnitude-Relation}
\label{sec:observationsComparison}
 To compare the predicted LFs and CMR with the observed ones we follow 
 \citet{Bouwens2014,Bouwens2015a} and adopt the same procedure to compute the LFs and $\langle \beta \rangle$
 from the synthetic galaxies SEDs. 
 
At each redshift, filters sample different ranges of the rest-frame galaxy SED. To be consistent with \citet{Bouwens2014},
we use the z850, Y105 and H160 filters for galaxies at $z \sim 5$, Y105 and H160 filters for $z \sim 6$, H160 and J125 filters for 
$z \sim 7$ and H160 and JH140 filters for $z \sim 8$\footnote{With
z850, Y105, J125, JH140 and H160 we refer to HST filters F850LP, F105W, F125W, F140W and F160W, respectively.}. 
 To compute the AB magnitude at each filter, we first define the pivot wavelength of a given filter $a$ with transmission $T_\lambda$ as, 
 
 \begin{equation}
 \lambda^{\rm a}_{\rm p} = \sqrt{\frac{\int_{-\infty}^{+\infty}  \lambda \, T_\lambda(\lambda) \, d\lambda}{\int_{-\infty}^{+\infty}  T_\lambda(\lambda)/\lambda \, d\lambda}}.
 \end{equation}
 \noindent
 Then, we compute the weighted filter flux $F_{\rm a}$ for a source at redshift $z$ with a given flux $f_\lambda$ 
 as, 
 \begin{equation}
 F_{\rm a} = \frac{\int_{-\infty}^{+\infty} \lambda \,  f(\lambda/(1+z))\, T(\lambda) \, d\lambda}
 				{\int_{-\infty}^{+\infty} \,  \lambda \, T(\lambda) \, d\lambda}.
 \end{equation}
 \noindent
 Finally, we define the absolute AB magnitude as,  
 \begin{equation}
	M^{\rm a}_{\rm AB} = -2.5 \log_{10} \left[\frac{F_{\rm a}}{\rm erg\,s^{-1}\,\MAA^{-1}}
	\left(\frac{\lambda^{\rm a}_{\rm p}}{\MAA}\right)^2 (1+z)^{-2} 
	\right] -97.78.
 \end{equation}
 \noindent
The $\beta$ slopes at $z \sim 5$ are computed as a least-square linear fit on the three filters, wheras at higher $z$  
we use the relation\footnote{To estimate the error introduced by this procedure, we compute the photometric $\beta$ for synthetic spectra with known $\beta$ in the
 range -3 and 0 and we find that the difference between the photometric determination and the input value is always less than 2\%.},
 \begin{equation}
   \beta = \frac{\log_{10} (F_{\rm a}/F_{\rm b}) }{\log_{10}( \lambda^{\rm a}_{\rm p}/\lambda^{\rm b}_{\rm p})}.
 \end{equation}
 \noindent
 The first step is slightly different from the method adopted by \citet{Bouwens2014}, where they use the effective wavelength assuming a power spectrum of 
 $\propto \lambda^{-2}$ instead of the pivot wavelength. 
Finally, to compare with the LFs computed by \citet{Bouwens2015a}, we evaluate the AB magnitude at 1600 \AA \, assuming a spectral slope given by the corresponding photometric $\beta$.

\begin{figure*}
\includegraphics[width=\textwidth]{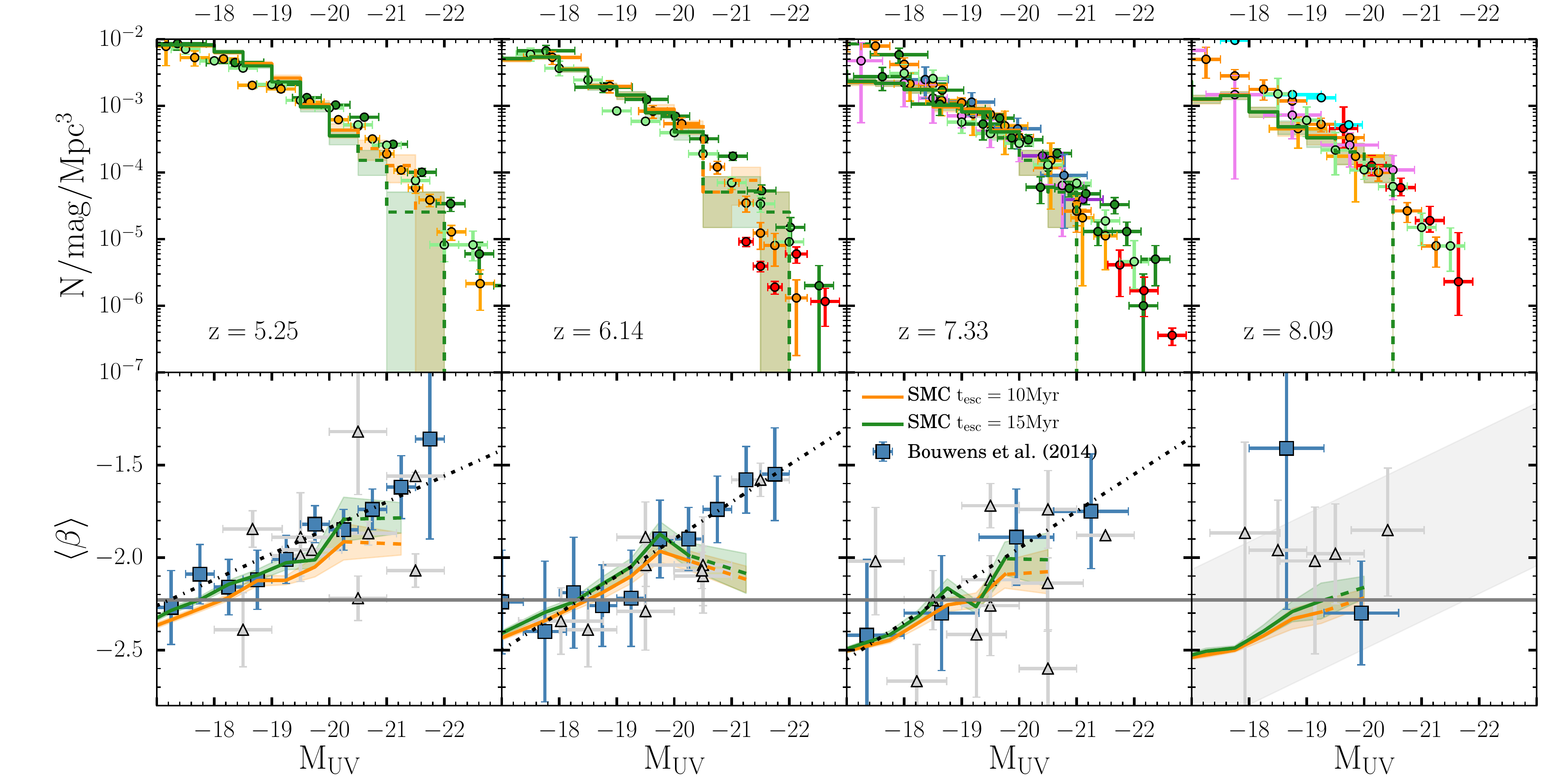}
\caption{Comparison between the predicted UV luminosity functions ({\it top panels}) and $\langle\beta\rangle$ slopes ({\it bottom panels})  with 
observations. Data are the same as in Fig. ~\ref{fig:combint}. In the bottom panels we have added observations from \citet{Wilkins2011}, \citet{Finkelstein2012}, 
\citet{Bouwens2012}, \citet{Dunlop2012, Dunlop2013}, and Duncan et al. (2014, all shown with grey data points). The black dot-dashed lines represent
the best-fit to the observations of Bouwens et al. (2014, blue data points) at $z \sim 5, 6,$ and $7$ and the shaded grey region is obtained extrapolating
the lower-$z$ slope and the best-fit intercept  at $z \sim 8$. 
The theoretical models adopt a SMC extinction curve and $t_{\rm esc} = 10$ (solid orange) and 15 Myr (solid green) with shaded regions representing
the Poissonian errors in each magnitude bin ({\it top panels}) and the standard errors on the mean values ({\it bottom panels}). 
Dashed lines indicate the luminosity range where less than 10 model galaxies are found in each magnitude bin (see text). The horizontal grey lines show
the value $\beta = -2.23$ adopted in the \citet{meurer1999} relation. 
A coloured version of this Figure is available online.}
\label{fig:lfbetaObs}
\end{figure*}

The results are plotted in Fig.~\ref{fig:lfbetaObs}. Model predictions at $5 \le z \le 8$ are obtained assuming the SMC extinction curve 
and $t_{\rm esc} = 10$ (orange curve) and 15 Myr (green curve). The shaded regions represent Poissonian errors associated to each
magnitude bin.  In the bottom panels we show the results of the systematic analysis by Bouwens et al. (2014, blue data points) and the corresponding best-fit relation
(black dot-dashed lines). For comparison, we also report data from \citet{Wilkins2011}, \citet{Finkelstein2012}, \citet{Bouwens2012}, \citet{Dunlop2012, Dunlop2013},
and \citet{Duncan2014}, all shown with grey data points. At $z \sim 8$, current observations of the $\beta$ slopes are highly uncertain, due to the small sizes of 
galaxy samples and photometric uncertainties introduced by the limited filter separation. The grey shaded region in the $z \sim 8$
bottom panel shows the CMR relation obtained extrapolating the lower-$z$ slope  and the best-fit intercept at $z \sim 8$
\citep{Bouwens2014}. 

Although both models appear to well reproduce the trend of an increasing reddening with luminosity observed by Bouwens et al. (2014) at $5 \le z \le 7$, at the brightest 
luminosities the statistics is too poor for a meaningful comparison. At each redshift, we identify a limiting luminosity above which the number of sources per magnitude bin is $< 10$,
and we illustrate the corresponding magnitude range with dashed lines. To better populate this luminosity range, a larger simulation volume would be required. In fact, 
at $5 \lesssim  z \lesssim 6$ the number of simulated galaxies with {\it intrinsic} $ -23 \leq M_{\rm UV} \leq - 20.5$ ranges between 40 and 65. These are the galaxies which suffer the largest dust extinction, with $\langle A_{\rm UV} \rangle \sim 1.7 \, (1.1)$ at $z \sim 5$ (6), and are observed at
$-21.3 \le M_{\rm UV} \leq -18.8$  ($-21.9 \le M_{\rm UV} \leq -19.4)$. Extrapolating these trends, we predict the brightest galaxies observed at 
$z \le 6$ with $M_{\rm UV} \lesssim -22$ to be massive ($M_{\rm star} > 5 \times 10^{10} M_\odot$) and dust-enriched ($M_{\rm dust} > 10^8 M_\odot$),
with typical $A_{\rm UV} > 1.7$, consistent with their relatively red observed colours, $\left\langle \beta \right\rangle \sim - 1.5$ (see Fig.~\ref{fig:Avsbeta}).

We conclude that while the ISM dust has a negligible effect on the galaxy UV LFs at $z \sim 7$ and $8$, it reduces the number of galaxies 
with $M_{\rm UV} \ge -18$ and $\ge -19$ at $z \sim 5$ and 6 to values in very good agreement with observations. 
The CMR
and its dependence on $z$ is sensitive to the extinction properties of the grains and to
the dust distribution in the ISM. In particular, {\it the observed trends suggest a steep extinction curve in the wavelength range $1500 \, \AA \le \lambda \le 3000 \, \AA$,
and that  stars with age $\leq 15$~Myr are embedded in their dense molecular natal clouds and their UV luminosity suffers a larger dust extinction.}

 When the grains are assumed to follow the MEC normalized to the SN extinction coefficient at $\lambda = 3000 \AA$ (see the curve shown in Fig.\ref{fig:extinctionCurve}), the 
 number of galaxies with $M_{\rm UV} \le -19$ at $z \sim 5$ is too small compared to the observed LF. Conversely, if the MEC is normalized to the SMC extinction
 coefficient at $\lambda = 3000 \AA$, the flatter slope at shorter wavelenghts reduces the predicted $\langle \beta \rangle$, at odds with observations. A better
 agreement is found if, following \citet{Gallerani2010}, the MEC is assumed to reflect a population of grains with intermediate properties
 between SN and SMC dust, and we adopt a normalization factor equal to $k_{\rm MEC} = (1- p) \, k_{\rm SMC} + p  \, k_{\rm SN}$ at $3000\,\AA$.
{\it Current observations do not allow to discriminate between the SMC and MEC models if $p \le 40\%$}. It is interesting to note that evidence for an SMC-like
extinction curve being preferred for galaxies at high-$z$ has been reported in many recent observational studies \citep{Tilvi2013, Oesch2013, Capak2015, Bouwens2016}.

Independently of the grain properties, the observed CMR requires dust evolution models 
in a 2-phase ISM, where SNe and AGB stars contribute to dust enrichment, 
dust grains grow their mass in dense molecular clouds, and are destroyed by SN shocks in the diffuse phase. 

 This conclusion is further strengthened by comparing our results with the recent studies by \citet{Shimizu2014}, Finkelstein et al. (2015, see in particular their Section 7), and \citet{Khakhaleva2016}. In these studies, the dust-to-gas mass ratio has been assumed to simply scale with the gas metallicity. In \citet{Shimizu2014}, they reproduce the 
observed UV-luminosity function and $\beta$ evolution with redshift at $z \ge 7$ by adjusting the dust-to-metal mass ratio, the effective radius of the dust distribution, and 
a parameter which controls the relative dust/star geometry. In the semi-analytical models that \citet{Finkelstein2015} compare with observations, a dust slab model
is adopted and the normalization of the dust optical depth is assumed to be $\propto {\rm exp}(-z/2)$ to obtain a reasonably good fit to the observed UV-LFs
at $z \geq 5$. They suggest that this scaling may be physically interpreted as due to an evolution of the dust-to-metal ratio or of the dust geometry. Our model allows
to predict the redshift and luminosity dependence of the dust optical depth, with the only free parameter being the residence time of young stars in molecular clouds.
Finally, using a 
dust radiative transfer model, \citet{Khakhaleva2016} reproduce the observed UV-LFs at $z \sim 6$ and 7, but their predictions are inconsistent with 
the data at $z \sim 8$ and lead to colour-magnitude relations that are shallower than observed. In their model dust is assumed to scale with metallicity and 
to be instantaneously sublimated in ionized regions. While the latter is certainly a reasonable assumption, it is not enough to capture the complex dynamical
interplay between dust formation and destruction in the different phases of the ISM, which is ultimately responsible for the observed evolution with redshift
and luminosity of dust extinction.

\subsection{Scatter in the $\beta - M_{\rm UV}$ and $M_{\rm star}-M_{\rm UV}$ relations}
\label{sec:scatter}

\begin{figure*}
\includegraphics[width=\textwidth]{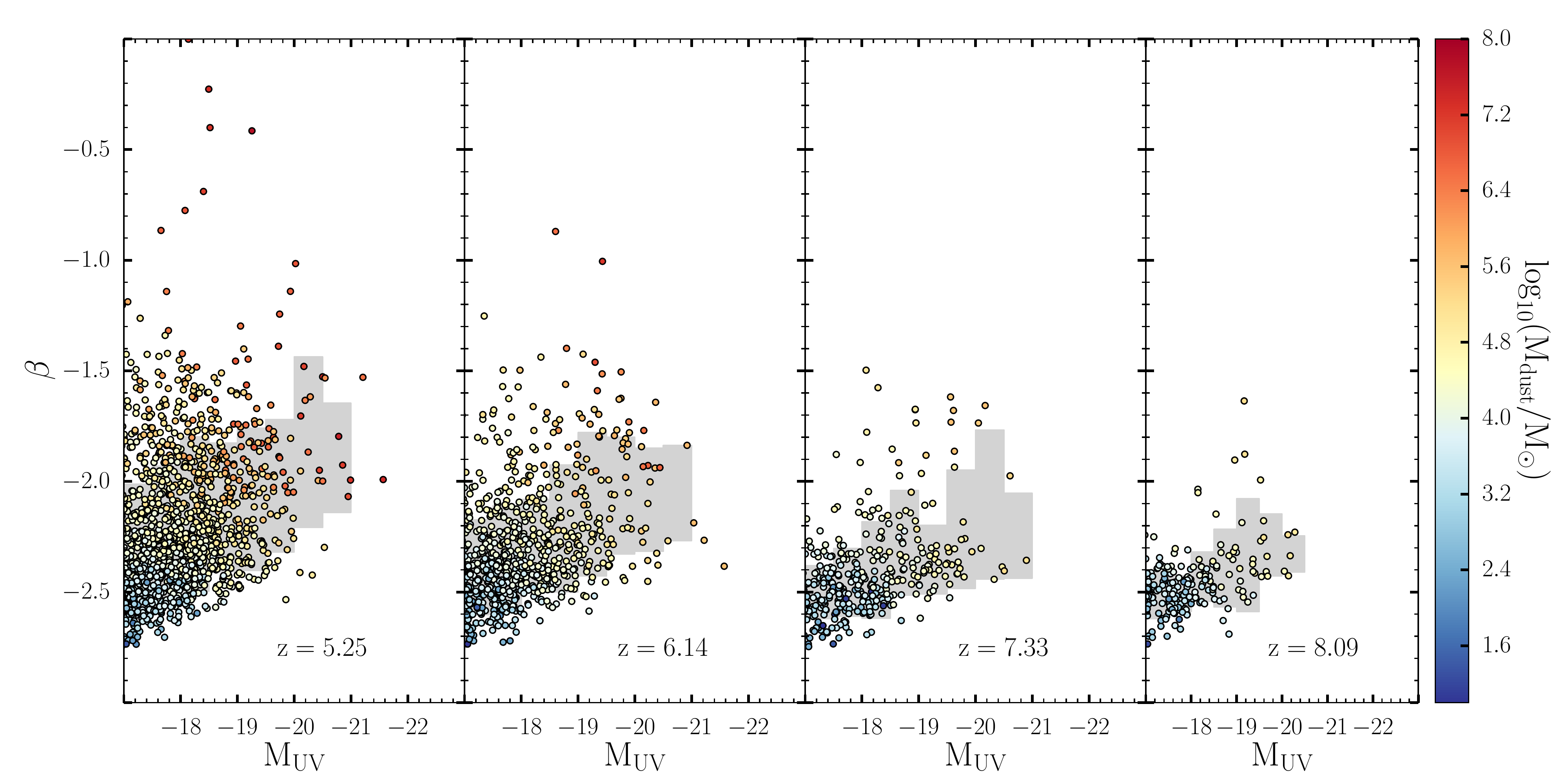}
\caption{Predicted $\beta$ slopes as a function of the UV magnitude at $z \sim 5, 6, 7$ and 8 (from left to right). Each data point
represents a galaxy and it is colour-coded according to the mass of dust in the ISM (colour scale on the right). We have assumed
the SMC extinction curve and $t_{\rm esc} = 15$~Myr. The grey shaded regions show the 1-$\sigma$ scatter around the CMR
shown in Fig.~\ref{fig:lfbetaObs}.
A coloured version of this Figure is available online.}
\label{fig:betascatter}
\end{figure*}
\begin{figure*}
\includegraphics[width=\textwidth]{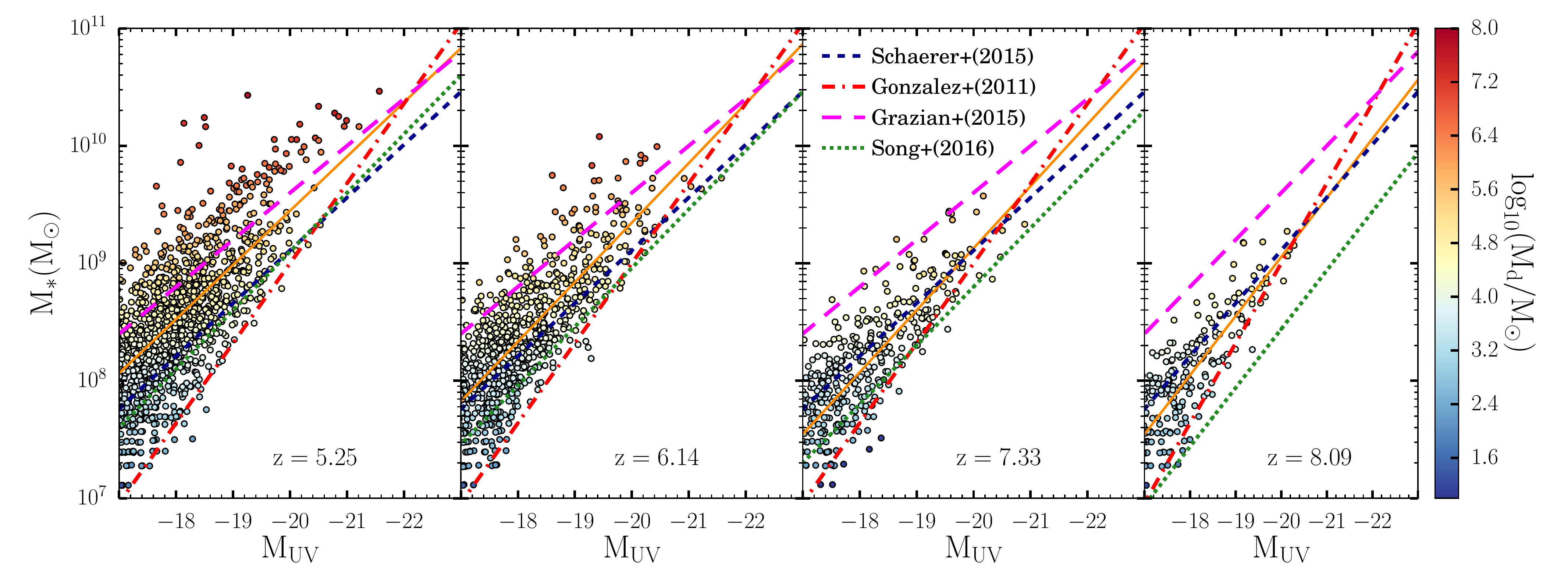}
\caption{Same as Fig.~\ref{fig:betascatter} but for the stellar mass. In each panel,
the solid line is the best-fit relation for the simulated galaxies and the other lines show the relations inferred from observational
data by \citet{Gonzalez2011}, \citet{Grazian2015}, \citet{Schaerer2015} and \citet{Song2016}.
A coloured version of this Figure is available online.}
\label{fig:mstarscatter}
\end{figure*}
\begin{figure*}
\includegraphics[width=.9\textwidth]{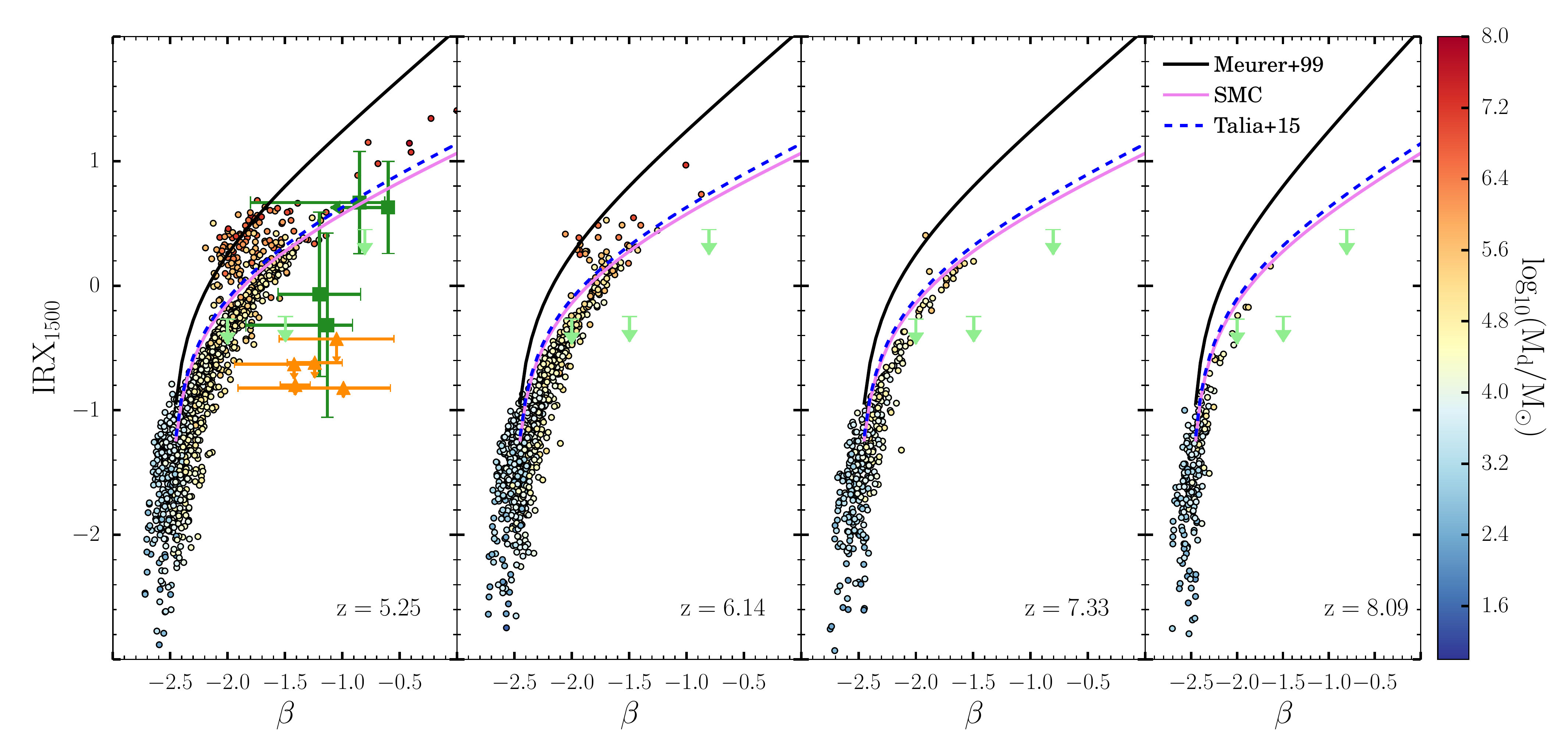}
\caption{The IR excess as a function of UV slope $\beta$ for the simulated galaxies at $z \sim 5, 6, 7$ and 8 (from left to right). Each data point
represents a galaxy and it is colour-coded according to the mass of dust in the ISM (colour scale on the right). We have assumed
the SMC extinction curve and $t_{\rm esc} = 15$~Myr. The black solid lines show the \citet{meurer1999} correlation with $\beta_0 = -2.5$, the blue dashed 
lines show the relation inferred by \citet{Talia2015}, and the magenta solid lines the relation predicted for the SMC extinction curve.
The data points at $z \sim 5$ represent the ALMA detected (green squares)
and ALMA non detected (orange triangles) sources reported by \citet{Capak2015}. The upper limits shown in all panels with light green triangles
are the results recently reported by \citet{Bouwens2016} for galaxies at $z \sim 4 -10$ (see text).
A coloured version of this Figure is available online.}
\label{fig:IRX}
\end{figure*}
\begin{figure*}
\includegraphics[width=\textwidth]{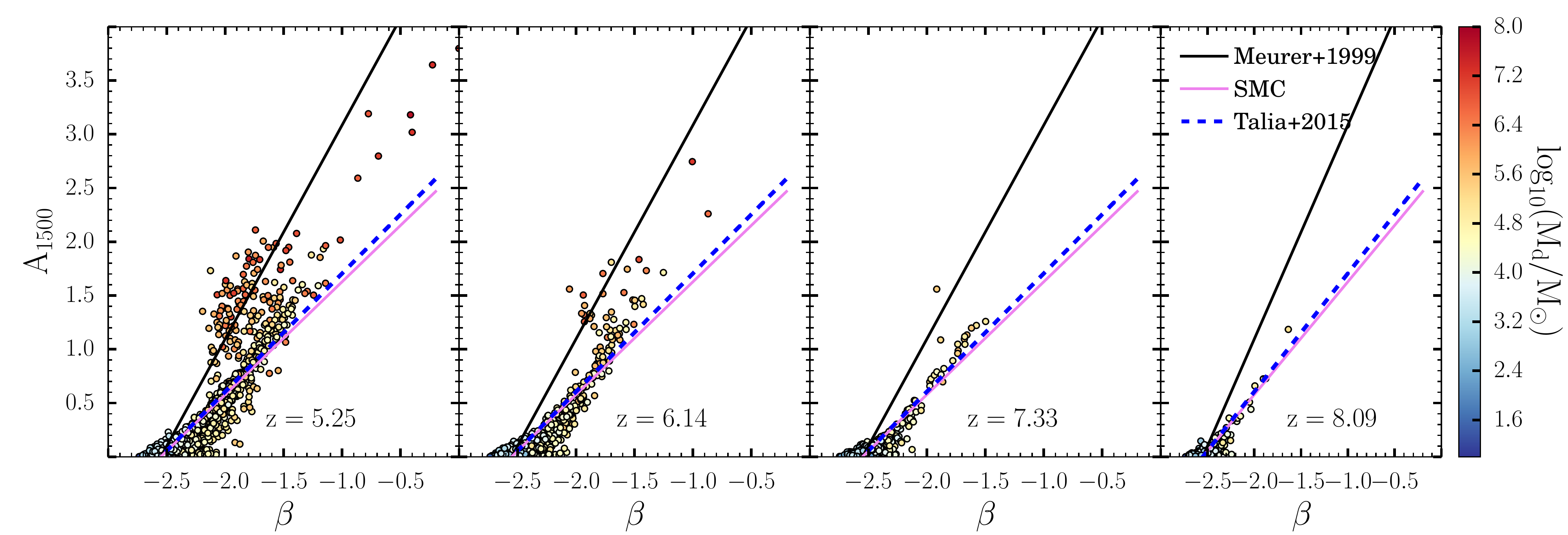}
\caption{Same as Fig.~\ref{fig:IRX} but for the dust attenuation factor at $1500\, \AA$.
A coloured version of this Figure is available online.}
\label{fig:Avsbeta}
\end{figure*}

High-$z$ galaxy samples show a considerable scatter in the measured $\beta$ slopes, even after accounting
for observational effects  \citep{Castellano2012,Bouwens2014, Rogers2014}. Studying the distribution of galaxy
colours at different redshifts can provide interesting indications on the origin and evolution of the CMR. 

Fig.~\ref{fig:betascatter}
shows the predicted distribution of galaxy colours at $z \sim 5, 6, 7, $ and 8 as a function of the UV magnitude,
assuming the SMC extinction curve and $t_{\rm esc} = 15$~Myr. Each data point represents an individual galaxy, colour-coded
depending on the mass of dust present in its ISM. The shaded regions show the $1-\sigma$ scatter around the CMR
shown in Fig.~\ref{fig:lfbetaObs} (green line) using the same UV magnitude bins. The amount of scatter in the colour distribution increases with
cosmic time, as a result of the progressively larger degree of dust enrichment. At each $z$, the scatter in the colour distribution
increases with luminosity, as the brightest galaxies are also more massive and dust enriched. This is consistent with
the analysis of \citet{Rogers2014} of a galaxy sample at $z \sim 5$, where they find an increasing
width of the colour distribution towards brighter galaxies. We find that there is a minimum value of $\beta$ that grows with UV luminosity
as a consequence of a minimum level of dust enrichment produced by stellar sources. This effect is independent of $z$ but the number
of galaxies at the bright end grows with time. At each $z$, galaxies with the reddest colours can have largely different luminosities:
for a given $\beta$ slope, the brightest galaxies are generally dustier. {\it At $z \lesssim 6$, sources with luminosities in the range
$-18 \le M_{\rm UV} \le -19$, where we have adequate statistics, appear to be a mix of intrinsically faint blue galaxies and of  
red objects which have suffered strong dust extinction. The latter population grows with cosmic time as a result of progressively 
more efficient grain-growth in their ISM}.

 Fig.~\ref{fig:mstarscatter} shows that the  population of dusty, UV-faint galaxies at $z \sim 5$ and 6 lie off the mean $M_{\rm star} - M_{\rm UV}$ relations
inferred from observations at comparable and higher-$z$ \citep{Gonzalez2011, Duncan2014, Grazian2015, Schaerer2015, Song2016}\footnote{Systematic
uncertainties associated with sample selection and stellar mass estimation lead to large discrepancies between different observational studies, even when
using the same data set \citep{Song2016}.
 At higher redshift, the
scatter is reduced and the simulated galaxies follow a tighter $M_{\rm star} - M_{\rm UV}$ relation. This may be an evolutionary effect, as $z \sim 7$ and 8 galaxies
have experienced limited dust enrichment. However, due to the
limited volume of our simulation, we can not exclude that massive, dusty, UV faint galaxies may have formed at these redshifts. 
Interestingly, there are observational evidences for massive, red galaxies at $z \sim 4 - 5$ \citep{Grazian2015,Song2016}.}
Using deep optical and infrared imaging provided
by HST, Spitzer and the VLT in the CANDELS-UDS, GOODS-South and HUDF, \citet{Grazian2015} show that the data at $3.5 < z < 4.5$ are consistent
with a constant mass-to-light ratio but with a considerable scatter. In particular, they find a population of relatively faint galaxies (with
$M_{\rm UV} \sim -18$) with masses $M_{\rm star} \sim 10^{11} M_\odot$, which can be comparable in number to UV bright galaxies with the same
stellar mass. Because of their red colours, these galaxies can not be selected by standard LBG criteria based on UV rest-frame colours. The difference
between the galaxy stellar mass function inferred from UV-selected star forming galaxies by  \citet{Gonzalez2011} and the mass function derived by
\citet{Duncan2014} and \citet{Grazian2015} has been interpreted as due to a growing contribution of massive dusty galaxies at $z \lesssim 5.5$.
While at higher redshifts there is better agreement, this may be due to a selection effect and the epoch of appearance of massive dusty galaxies may
require future deep infrared surveys \citep{Grazian2015}.

\subsection{The IR excess}
\label{sec:IRprop}

Dust attenuation of star forming galaxies at high redshift is commonly evaluated using methods based 
on the observed correlation between the spectral slope $\beta$ and the infrared excess, IRX \citep{meurer1999}.
The latter is defined as the ratio between the IR to UV fluxes (at $\lambda = 1600 \, \AA$), ${\rm IRX }= {\rm Log} \, F_{\rm IR}/F_{\rm 1600}$
and it is a measure of dust absorption. Hence the IRX-$\beta$ relation shows that dust absorption is correlated to UV reddening and provides
a powerful tool to reconstruct the unattenuated UV flux when only UV rest-frame data is available. The original idea was proposed by \citet{meurer1999}
using a sample of local starbursts from which the following relations were derived:
\begin{equation}
{\rm IRX} = {\rm Log} (10^{0.4 A_{1600}} - 1) + 0.076 \pm 0.044,
\end{equation}
\noindent
and
\begin{equation}
A_{1600} = 4.43 + 1.99 \, \beta,
\end{equation}
where $A_{1600} = 1.086 \, \tau_{1600}$ is the dust attenuation at $1600 \, \AA$ and the dispersion on the fit was 0.55 mag in $A_{1600}$ and 
0.28 on $\beta$ \citep{meurer1999}. The relation obviously depends on the intrinsic spectrum of the
sources and on the extinction curve. The zero point of the relation implies that the intrinsic spectral slope of the sources is assumed to be $\beta_0 = -2.23$.
This value has been reported as a grey solid line in Figs.~\ref{fig:comboLfBeta} - \ref{fig:lfbetaObs} to show that all
the simulated galaxies at $5 \le z \le 8$ have bluer intrinsic colours. Indeed, modifications of the original relation to account for the lower metallicities
and younger ages of galaxies at high redshift have been proposed in the literature. Using a small  sample of galaxies at $z \sim 2.8 - 3$
with deep IR observations and measured spectroscopic metallicities from the CANDELS+HUGS survey, \citet{Castellano2014} derived the relation
$A_{\rm 1600} =  5.32 +1.99 \, \beta$. This implies a value of the intrinsic slope $\beta_0 = -2.67$, consistent with the sub-solar metallicities and young ages inferred
for their sample galaxies. In Figs.~\ref{fig:IRX} and \ref{fig:Avsbeta} we show the \citet{meurer1999} relation modified assuming a value of $\beta_0 = -2.5$, the mean intrinsic colours of the
simulated galaxies (solid black lines). In the same figures, we also show the much flatter relations derived by \citet{Talia2015} using the UV spectra of a 
sample of 62 IR-selected galaxies at $1 < z < 3$ (blue dashed lines), which is more consistent with the relation inferred by \citet{Pettini1998} for the SMC extinction
curve (solid magenta lines). The inferred IRX - $\beta$ is known to depend on the galaxy sample selection method. 
While UV and optically selected samples distribute systematically lower than starbursts on the IRX - $\beta$ plane \citep{Cortese2006, Boissier2007}, 
two different distributions are found in IR-selected samples. Luminous and Ultraluminous IR galaxies distribute above the \citet{meurer1999} relation \citep{Goldader2002,
Takeuchi2010, Howell2010, Reddy2010, Overzier2011, Casey2014, Forrest2016}, quiescent star forming galaxies distribute below it \citep{Takeuchi2010, Buat2012, Talia2015}.

The figures also show the simulated galaxies at $5 \le z \le 8$, colour-coded depending on the level of dust enrichment. We have assumed the SMC extinction curve and $t_{\rm esc} = 15$~Myr. For each galaxy, we compute the IRX at $1500 \, \AA$
assuming that all the absorbed UV radiation is re-emitted in the IR. We account also for the contribution of resonantly scattered Lyman-$\alpha$ photons, 
which is estimated to be 7\% of the UV radiation \citep{Khakhaleva2016}. 

At $z \sim 7$ and 8, we find that all the simulated galaxies  are characterized by an IRX considerably smaller than that predicted by the
\citet{meurer1999} relation, and are consistent with that predicted for the SMC and the one derived by 
\citet{Talia2015}. However, a second population of dusty galaxies appears at $z \lesssim 6$, which progressively shifts towards the \citet{meurer1999} relation, although
with a large scatter. This is the same population that lies off the stellar mass - UV luminosity relation shown in Fig.~\ref{fig:mstarscatter} and that dominates the scatter 
in the colour distribution shown in Fig.~\ref{fig:betascatter}. {\it Our analysis suggests that lower stellar mass and less chemically mature galaxies
at high-$z$ are characterized by smaller IRX and $A_{\rm 1500}$ than implied by the \citet{meurer1999}
relation for galaxies with the same colours.} Their ISM dust is mostly contributed by stellar sources and their dust attenuation is smaller, consistent with what has been found for young ($< 100$~Myr) LBGs at $z \sim 3$ by \citet{Siana2009} and \citet{Reddy2010},  and more recently by \citep{Bouwens2016} using ALMA 1.22 mm-continuum observations 
of a 1 arcmin$^2$ region in the Hubble Ultra Deep Field. 
However, we find that {\it massive and more chemically evolved galaxies, where grain growth in dense gas increases the mass of ISM dust, introduce a considerable 
scatter in the IRX at a given UV continuum slope}. At $z \sim 5.25$, galaxies with $-2 < \beta < -1.5$ can have IRX in the range 0.3 - 4 and it is very hard to infer the proper
dust attenuation factor from the UV slope alone (see Fig.\ref{fig:Avsbeta}). 

In Fig.~\ref{fig:IRX} we also show the IR excess of $z \sim 5.1 - 5.7$ galaxies inferred by \citet{Capak2015}, which have been argued to be significantly more 
dust-poor and less IR-luminous than lower $z$ galaxies with similar UV colours.
To be consistent with the data points shown in Fig.\ref{fig:dustm}, we have computed the IRX values of the \citet{Capak2015} sources, from their measured (or upper limits) 
$\rm 158\; \mu m$ flux, adopting a modified 
black body spectrum with emissivity index $\beta = 1.5$ \citep{Ota2014} and a dust temperature $T_{\rm d} = 35$~K, as described in Section 3.1. This yields values 
of the FIR emissivities that are $25 \%$ larger, but consistent within the errors, with the ones reported by (\citealt{Capak2015}, see their Table 5). We find that simulated galaxies which follow the SMC and the \citet{Talia2015} correlations at the same $z$
are marginally compatible with the IRX of the ALMA detected sources (green squares), given the large uncertainties on their $\beta$ slopes. However, the simulated galaxies with
IRX compatible with the upper limits inferred for the ALMA undetected sources (orange triangles) have significantly bluer colours, consistent with their low dust content. 
Our study confirms that  it is difficult to explain the low IRX of the \citet{Capak2015} sources, unless their $\beta$ slopes have been overestimated or the dust temperature
(hence the FIR flux) has been underestimated. A similar conclusion applies to the recent results reported by \citet{Bouwens2016} using stacked constraints on the IRX for a
sample of $z \sim 4 - 10$ galaxies of the HUDF obtained with deep 1.2 mm-continuum observations (see the upper limits in Fig.~\ref{fig:IRX}). A more detailed analysis of these
latest findings is deferred to a future study.

\section{Conclusions}
\label{sec:conclusions}

The main goal of the present study is to provide a consistent framework to interpret the observed evolution of the UV LFs and galaxy colours over the 
redshift range $5 \leq z \leq 8$. To this aim, we have used a semi-numerical approach to post-process the output of a cosmological simulation 
with a chemical evolution model with dust. Our approach allows us to follow dust enrichment by stellar sources (SNe and AGB stars), dust
destruction in the diffuse gas by SN shocks, and grain growth in dense molecular clouds. The model has already been applied by \citet{mancini2015}
to interpret current observational constraints on the dust mass inferred from ALMA and Plateau de Bure observations of normal star forming
galaxies at $z > 6$. Here we extend the analysis to investigate how dust properties affect the UV LFs and galaxy colours at high-$z$. 
Our main findings can be summarized as follows:

\begin{itemize}
\item The comparison between model predictions and observations at $5 \lesssim z \lesssim 8$ shows that, while the ISM dust has a negligible effect on the galaxy UV LFs at $z \sim 7$ and $8$, it reduces the number of galaxies with $M_{\rm UV} \ge -18$ and $\ge -19$ at $z \sim 5$ and 6 to values in very good agreement with observations. 
The observed  CMR and its dependence on $z$ suggest a steep extinction curve in the wavelength range $1500 \, \AA \le \lambda \le 3000 \, \AA$,
and that  stars with age $\leq 15$~Myr are embedded in their dense molecular natal clouds and their UV luminosity suffers a larger dust extinction. 
\item The scatter in the colour distribution around the mean CMR increases with luminosity and cosmic time, consistent with observations.  At $z \lesssim 6$,
galaxies with $-19 \leq M_{\rm UV} \leq -18$ (where we have adequate statistics, given our simulation volume and resolution) are a mix of intrinsically faint
blue galaxies and of red objects which have suffered strong dust extinction. The latter population grows with time, as a result of more efficient grain-growth
in their ISM, and lie off the mean $M_{\rm star} - M_{\rm UV}$ relation inferred from observations at $z \sim 5,$ and 6. This is supported by the recent evidence for
a population of massive UV-faint objects that makes a non negligible contribution to the stellar mass function at  $z \lesssim 5.5$ \citep{Grazian2015}.
\item By analyizing the properties of the simulated galaxies in the IRX - $\beta$ plane, we find that young, less massive galaxies, where the ISM dust is mostly
contributed by stellar sources, follow a relation which is much flatter than the commonly adopted \citet{meurer1999} relation, consistent with their
steep extinction curve. Massive dusty galaxies, which have experienced efficient grain growth in their ISM, introduce a considerable scatter in the
IRX at a given UV continuum slope, slowly shifting towards the \citet{meurer1999} relation at $z \lesssim 6$. 
\item At $z \sim 7$ and 8, dust attenuation factors are better estimated assuming a flatter IRX - $\beta$ relation, such as the 
one recently inferred by \citet{Talia2015} or predicted for the SMC curve \citep{Pettini1998}. At $z \lesssim 6$, it is very hard to infer the proper dust 
attenuation from the UV slope alone, as galaxies with $-2 < \beta < -1.5$ can have vastly different IRX.  
\end{itemize}

Our analysis suggests that {\it the total star formation rate density at high-$z$ may be overestimated if dust attenuation factors are derived 
using the \citet{meurer1999} relation, and that more realistic dust correction for young galaxies, which have not yet experienced major dust enrichment,
can be derived from their UV colours using a flatter IRX - $\beta$ relation, such as the one inferred by \citet{Talia2015} or implied by the SMC curve. 
However, once grain growth starts to dominate dust enrichment, a population of massive, dusty, and UV faint galaxies appears at $z \lesssim 6$. 
These galaxies increase the scatter in the $\beta - M_{\rm UV}$, $M_{\rm star} - \beta$ and IRX - $\beta$ planes and slowly shift towards the \citet{meurer1999}
relation. }

We do not find dusty, massive, UV-faint galaxies at $z \sim 7$  and 8, but we can not exclude this to be an effect of the limited volume of our simulation. 
In fact, at the
same redshifts we also underpredict the bright-end of the observed UV LFs, even assuming no dust extinction.
Despite these limitations, our study shows that current high-$z$ observations on the evolution of galaxy colours already provide important constraints
on the nature of dust and on its complex evolution and spatial distribution in the interstellar medium. The next mandatory step is to incorporate these
processes directly into numerical simulations.

\section*{Acknowledgments}
We thank Marco Castellano for his kind clarifications and Andrea Ferrara, Seiji Fujimoto, Hiroyuki Hirashita, Andrea Pallottini and Livia Vallini 
for insightful comments.
R. Schneider, M. Mancini and L. Graziani acknowledge the hospitality of the KITP, where this work
was completed. 
The research leading to these results has received funding from the European Research Council under the European 
Union Seventh Framework Programme 
(FP/2007-2013) / ERC Grant Agreement n. 306476, and by the National Science Foundation under Grant No. NSF PHY11-25915.

\bibliography{bibliography}

\begin{thebibliography}{}
\makeatletter
\relax
\def\mn@urlcharsother{\let\do\@makeother \do\$\do\&\do\#\do\^\do\_\do\%\do\~}
\def\mn@doi{\begingroup\mn@urlcharsother \@ifnextchar [ {\mn@doi@}
  {\mn@doi@[]}}
\def\mn@doi@[#1]#2{\def\@tempa{#1}\ifx\@tempa\@empty \href
  {http://dx.doi.org/#2} {doi:#2}\else \href {http://dx.doi.org/#2} {#1}\fi
  \endgroup}
\def\mn@eprint#1#2{\mn@eprint@#1:#2::\@nil}
\def\mn@eprint@arXiv#1{\href {http://arxiv.org/abs/#1} {{\tt arXiv:#1}}}
\def\mn@eprint@dblp#1{\href {http://dblp.uni-trier.de/rec/bibtex/#1.xml}
  {dblp:#1}}
\def\mn@eprint@#1:#2:#3:#4\@nil{\def\@tempa {#1}\def\@tempb {#2}\def\@tempc
  {#3}\ifx \@tempc \@empty \let \@tempc \@tempb \let \@tempb \@tempa \fi \ifx
  \@tempb \@empty \def\@tempb {arXiv}\fi \@ifundefined
  {mn@eprint@\@tempb}{\@tempb:\@tempc}{\expandafter \expandafter \csname
  mn@eprint@\@tempb\endcsname \expandafter{\@tempc}}}

\bibitem[\protect\citeauthoryear{{Asano}, {Takeuchi}, {Hirashita}  \&
  {Inoue}}{{Asano} et~al.}{2013}]{Asano2013}
{Asano} R.~S.,  {Takeuchi} T.~T.,  {Hirashita} H.,   {Inoue} A.~K.,  2013,
  \mn@doi [Earth, Planets, and Space] {10.5047/eps.2012.04.014}, \href
  {http://adsabs.harvard.edu/abs/2013EP%26S...65..213A} {65, 213}

\bibitem[\protect\citeauthoryear{{Atek} et~al.,}{{Atek}
  et~al.}{2015}]{Atek2015}
{Atek} H.,  et~al., 2015, \mn@doi [\apj] {10.1088/0004-637X/800/1/18}, \href
  {http://adsabs.harvard.edu/abs/2015ApJ...800...18A} {800, 18}

\bibitem[\protect\citeauthoryear{{Bianchi} \& {Schneider}}{{Bianchi} \&
  {Schneider}}{2007}]{Bianchi2007}
{Bianchi} S.,  {Schneider} R.,  2007, \mn@doi [\mnras]
  {10.1111/j.1365-2966.2007.11829.x}, \href
  {http://ads.ari.uni-heidelberg.de/abs/2007MNRAS.378..973B} {378, 973}

\bibitem[\protect\citeauthoryear{{Bocchio}, {Jones}  \& {Slavin}}{{Bocchio}
  et~al.}{2014}]{Bocchio2014}
{Bocchio} M.,  {Jones} A.~P.,   {Slavin} J.~D.,  2014, \mn@doi [\aap]
  {10.1051/0004-6361/201424368}, \href
  {http://adsabs.harvard.edu/abs/2014A%26A...570A..32B} {570, A32}

\bibitem[\protect\citeauthoryear{{Bocchio}, {Marassi}, {Schneider}, {Bianchi},
  {Limongi}  \& {Chieffi}}{{Bocchio} et~al.}{2016}]{Bocchio2016}
{Bocchio} M.,  {Marassi} S.,  {Schneider} R.,  {Bianchi} S.,  {Limongi} M.,
  {Chieffi} A.,  2016, \aap, 587, A157

\bibitem[\protect\citeauthoryear{{Boissier} et~al.,}{{Boissier}
  et~al.}{2007}]{Boissier2007}
{Boissier} S.,  et~al., 2007, \apjs, 173, 524

\bibitem[\protect\citeauthoryear{{Bouwens} et~al.,}{{Bouwens}
  et~al.}{2011}]{Bouwens2011}
{Bouwens} R.~J.,  et~al., 2011, \mn@doi [\apj] {10.1088/0004-637X/737/2/90},
  \href {http://adsabs.harvard.edu/abs/2011ApJ...737...90B} {737, 90}

\bibitem[\protect\citeauthoryear{{Bouwens} et~al.,}{{Bouwens}
  et~al.}{2012}]{Bouwens2012}
{Bouwens} R.~J.,  et~al., 2012, \mn@doi [\apj] {10.1088/0004-637X/754/2/83},
  \href {http://ads.ari.uni-heidelberg.de/abs/2012ApJ...754...83B} {754, 83}

\bibitem[\protect\citeauthoryear{{Bouwens} et~al.,}{{Bouwens}
  et~al.}{2014}]{Bouwens2014}
{Bouwens} R.~J.,  et~al., 2014, \mn@doi [\apj] {10.1088/0004-637X/793/2/115},
  \href {http://adsabs.harvard.edu/abs/2014ApJ...793..115B} {793, 115}

\bibitem[\protect\citeauthoryear{{Bouwens} et~al.,}{{Bouwens}
  et~al.}{2015}]{Bouwens2015a}
{Bouwens} R.~J.,  et~al., 2015, \mn@doi [\apj] {10.1088/0004-637X/803/1/34},
  \href {http://adsabs.harvard.edu/abs/2015ApJ...803...34B} {803, 34}

\bibitem[\protect\citeauthoryear{{Bouwens} et~al.,}{{Bouwens}
  et~al.}{2016}]{Bouwens2016}
{Bouwens} R.,  et~al., 2016, preprint, \href
  {http://adsabs.harvard.edu/abs/2016arXiv160605280B} {} (\mn@eprint {arXiv}
  {1606.05280})

\bibitem[\protect\citeauthoryear{{Bowler} et~al.,}{{Bowler}
  et~al.}{2014}]{Bowler2014}
{Bowler} R.~A.~A.,  et~al., 2014, \mn@doi [\mnras] {10.1093/mnras/stu449},
  \href {http://adsabs.harvard.edu/abs/2014MNRAS.440.2810B} {440, 2810}

\bibitem[\protect\citeauthoryear{{Bowler} et~al.,}{{Bowler}
  et~al.}{2015}]{Bowler2015}
{Bowler} R.~A.~A.,  et~al., 2015, \mn@doi [\mnras] {10.1093/mnras/stv1403},
  \href {http://adsabs.harvard.edu/abs/2015MNRAS.452.1817B} {452, 1817}

\bibitem[\protect\citeauthoryear{{Buat} et~al.,}{{Buat}
  et~al.}{2012}]{Buat2012}
{Buat} V.,  et~al., 2012, \aap, 545, A141

\bibitem[\protect\citeauthoryear{{Calzetti}, {Kinney}  \&
  {Storchi-Bergmann}}{{Calzetti} et~al.}{1994}]{Calzetti1994}
{Calzetti} D.,  {Kinney} A.~L.,   {Storchi-Bergmann} T.,  1994, \mn@doi [\apj]
  {10.1086/174346}, \href
  {http://ads.ari.uni-heidelberg.de/abs/1994ApJ...429..582C} {429, 582}

\bibitem[\protect\citeauthoryear{{Calzetti}, {Armus}, {Bohlin}, {Kinney},
  {Koornneef}  \& {Storchi-Bergmann}}{{Calzetti} et~al.}{2000}]{Calzetti2000}
{Calzetti} D.,  {Armus} L.,  {Bohlin} R.~C.,  {Kinney} A.~L.,  {Koornneef} J.,
   {Storchi-Bergmann} T.,  2000, \mn@doi [\apj] {10.1086/308692}, \href
  {http://ads.ari.uni-heidelberg.de/abs/2000ApJ...533..682C} {533, 682}

\bibitem[\protect\citeauthoryear{{Capak} et~al.,}{{Capak}
  et~al.}{2015}]{Capak2015}
{Capak} P.~L.,  et~al., 2015, \nat, 522

\bibitem[\protect\citeauthoryear{{Cardelli}, {Clayton}  \& {Mathis}}{{Cardelli}
  et~al.}{1989}]{Cardelli1989}
{Cardelli} J.~A.,  {Clayton} G.~C.,   {Mathis} J.~S.,  1989, \apj, 345, 245

\bibitem[\protect\citeauthoryear{{Casey} et~al.,}{{Casey}
  et~al.}{2014}]{Casey2014}
{Casey} C.~M.,  et~al., 2014, \apj, 796, 95

\bibitem[\protect\citeauthoryear{{Castellano} et~al.,}{{Castellano}
  et~al.}{2010}]{castellano2010}
{Castellano} M.,  et~al., 2010, \mn@doi [\aap] {10.1051/0004-6361/201015195},
  \href {http://adsabs.harvard.edu/abs/2010A%26A...524A..28C} {524, A28}

\bibitem[\protect\citeauthoryear{{Castellano} et~al.,}{{Castellano}
  et~al.}{2012}]{Castellano2012}
{Castellano} M.,  et~al., 2012, \mn@doi [\aap] {10.1051/0004-6361/201118050},
  \href {http://adsabs.harvard.edu/abs/2012A%26A...540A..39C} {540, A39}

\bibitem[\protect\citeauthoryear{{Castellano} et~al.,}{{Castellano}
  et~al.}{2014}]{Castellano2014}
{Castellano} M.,  et~al., 2014, \aap, 566, A19

\bibitem[\protect\citeauthoryear{{Charlot} \& {Fall}}{{Charlot} \&
  {Fall}}{2000}]{Charlot2000}
{Charlot} S.,  {Fall} S.~M.,  2000, \apj, 539, 718

\bibitem[\protect\citeauthoryear{{Cherchneff} \& {Dwek}}{{Cherchneff} \&
  {Dwek}}{2009}]{Cherchneff2009}
{Cherchneff} I.,  {Dwek} E.,  2009, \mn@doi [\apj]
  {10.1088/0004-637X/703/1/642}, \href
  {http://adsabs.harvard.edu/abs/2009ApJ...703..642C} {703, 642}

\bibitem[\protect\citeauthoryear{{Cherchneff} \& {Lilly}}{{Cherchneff} \&
  {Lilly}}{2008}]{Cherchneff2008}
{Cherchneff} I.,  {Lilly} S.,  2008, \mn@doi [\apjl] {10.1086/591906}, \href
  {http://adsabs.harvard.edu/abs/2008ApJ...683L.123C} {683, L123}

\bibitem[\protect\citeauthoryear{{Cortese} et~al.,}{{Cortese}
  et~al.}{2006}]{Cortese2006}
{Cortese} L.,  et~al., 2006, \apj, 637, 242

\bibitem[\protect\citeauthoryear{{Dayal} \& {Ferrara}}{{Dayal} \&
  {Ferrara}}{2012}]{Dayal2012}
{Dayal} P.,  {Ferrara} A.,  2012, \mn@doi [\mnras]
  {10.1111/j.1365-2966.2012.20486.x}, \href
  {http://ads.ari.uni-heidelberg.de/abs/2012MNRAS.421.2568D} {421, 2568}

\bibitem[\protect\citeauthoryear{{Dayal}, {Ferrara}  \& {Saro}}{{Dayal}
  et~al.}{2010}]{Dayal2010a}
{Dayal} P.,  {Ferrara} A.,   {Saro} A.,  2010, \mnras, 402, 1449

\bibitem[\protect\citeauthoryear{{Dayal}, {Dunlop}, {Maio}  \&
  {Ciardi}}{{Dayal} et~al.}{2013}]{dayal2013}
{Dayal} P.,  {Dunlop} J.~S.,  {Maio} U.,   {Ciardi} B.,  2013, \mn@doi [\mnras]
  {10.1093/mnras/stt1108}, \href
  {http://adsabs.harvard.edu/abs/2013MNRAS.434.1486D} {434, 1486}

\bibitem[\protect\citeauthoryear{{Dayal}, {Ferrara}, {Dunlop}  \&
  {Pacucci}}{{Dayal} et~al.}{2014}]{Dayal2014}
{Dayal} P.,  {Ferrara} A.,  {Dunlop} J.~S.,   {Pacucci} F.,  2014, \mn@doi
  [\mnras] {10.1093/mnras/stu1848}, \href
  {http://ads.ari.uni-heidelberg.de/abs/2014MNRAS.445.2545D} {445, 2545}

\bibitem[\protect\citeauthoryear{{Di Criscienzo} et~al.,}{{Di Criscienzo}
  et~al.}{2013}]{Dicriscienzo2013}
{Di Criscienzo} M.,  et~al., 2013, \mn@doi [\mnras] {10.1093/mnras/stt732},
  \href {http://adsabs.harvard.edu/abs/2013MNRAS.433..313D} {433, 313}

\bibitem[\protect\citeauthoryear{{Draine}}{{Draine}}{2011}]{Draine2011}
{Draine} B.~T.,  2011, Physics of the Interstellar and Intergalactic Medium.
Princeton Univ. Press

\bibitem[\protect\citeauthoryear{{Duncan} et~al.,}{{Duncan}
  et~al.}{2014}]{Duncan2014}
{Duncan} K.,  et~al., 2014, \mnras, 444, 2960

\bibitem[\protect\citeauthoryear{{Dunlop}, {McLure}, {Robertson}, {Ellis},
  {Stark}, {Cirasuolo}  \& {de Ravel}}{{Dunlop} et~al.}{2012}]{Dunlop2012}
{Dunlop} J.~S.,  {McLure} R.~J.,  {Robertson} B.~E.,  {Ellis} R.~S.,  {Stark}
  D.~P.,  {Cirasuolo} M.,   {de Ravel} L.,  2012, \mnras, 420, 901

\bibitem[\protect\citeauthoryear{{Dunlop} et~al.,}{{Dunlop}
  et~al.}{2013}]{Dunlop2013}
{Dunlop} J.~S.,  et~al., 2013, \mnras, 432, 3520

\bibitem[\protect\citeauthoryear{{Ferrara}, {Pettini}  \&
  {Shchekinov}}{{Ferrara} et~al.}{2000}]{Ferrara2000}
{Ferrara} A.,  {Pettini} M.,   {Shchekinov} Y.,  2000, \mn@doi [\mnras]
  {10.1046/j.1365-8711.2000.03857.x}, \href
  {http://ads.ari.uni-heidelberg.de/abs/2000MNRAS.319..539F} {319, 539}

\bibitem[\protect\citeauthoryear{{Ferrarotti} \& {Gail}}{{Ferrarotti} \&
  {Gail}}{2006}]{Ferrarotti2006}
{Ferrarotti} A.~S.,  {Gail} H.-P.,  2006, \mn@doi [\aap]
  {10.1051/0004-6361:20041198}, \href
  {http://adsabs.harvard.edu/abs/2006A%26A...447..553F} {447, 553}

\bibitem[\protect\citeauthoryear{{Finkelstein} et~al.,}{{Finkelstein}
  et~al.}{2012}]{Finkelstein2012}
{Finkelstein} S.~L.,  et~al., 2012, \mn@doi [\apj]
  {10.1088/0004-637X/756/2/164}, \href
  {http://adsabs.harvard.edu/abs/2012ApJ...756..164F} {756, 164}

\bibitem[\protect\citeauthoryear{{Finkelstein} et~al.,}{{Finkelstein}
  et~al.}{2015a}]{Finkelstein2015}
{Finkelstein} S.~L.,  et~al., 2015a, \apj, 810, 71

\bibitem[\protect\citeauthoryear{{Finkelstein} et~al.,}{{Finkelstein}
  et~al.}{2015b}]{Finkelstein2015b}
{Finkelstein} S.~L.,  et~al., 2015b, \mn@doi [\apj]
  {10.1088/0004-637X/814/2/95}, \href
  {http://adsabs.harvard.edu/abs/2015ApJ...814...95F} {814, 95}

\bibitem[\protect\citeauthoryear{{Forero-Romero}, {Yepes}, {Gottl{\"o}ber},
  {Knollmann}, {Khalatyan}, {Cuesta}  \& {Prada}}{{Forero-Romero}
  et~al.}{2010}]{ForeroRomero2010}
{Forero-Romero} J.~E.,  {Yepes} G.,  {Gottl{\"o}ber} S.,  {Knollmann} S.~R.,
  {Khalatyan} A.,  {Cuesta} A.~J.,   {Prada} F.,  2010, \mnras, 403, L31

\bibitem[\protect\citeauthoryear{{Forrest} et~al.,}{{Forrest}
  et~al.}{2016}]{Forrest2016}
{Forrest} B.,  et~al., 2016, \apjl, 818, L26

\bibitem[\protect\citeauthoryear{{Gallerani} et~al.,}{{Gallerani}
  et~al.}{2010}]{Gallerani2010}
{Gallerani} S.,  et~al., 2010, \aap, 523, A85

\bibitem[\protect\citeauthoryear{{Goldader}, {Meurer}, {Heckman}, {Seibert},
  {Sanders}, {Calzetti}  \& {Steidel}}{{Goldader} et~al.}{2002}]{Goldader2002}
{Goldader} J.~D.,  {Meurer} G.,  {Heckman} T.~M.,  {Seibert} M.,  {Sanders}
  D.~B.,  {Calzetti} D.,   {Steidel} C.~C.,  2002, \apj, 568, 651

\bibitem[\protect\citeauthoryear{{Gonzalez-Perez}, {Lacey}, {Baugh}, {Frenk}
  \& {Wilkins}}{{Gonzalez-Perez} et~al.}{2013}]{GonzalezPerez2013}
{Gonzalez-Perez} V.,  {Lacey} C.~G.,  {Baugh} C.~M.,  {Frenk} C.~S.,
  {Wilkins} S.~M.,  2013, \mn@doi [\mnras] {10.1093/mnras/sts446}, \href
  {http://adsabs.harvard.edu/abs/2013MNRAS.429.1609G} {429, 1609}

\bibitem[\protect\citeauthoryear{{Gonz{\'a}lez}, {Labb{\'e}}, {Bouwens},
  {Illingworth}, {Franx}  \& {Kriek}}{{Gonz{\'a}lez}
  et~al.}{2011}]{Gonzalez2011}
{Gonz{\'a}lez} V.,  {Labb{\'e}} I.,  {Bouwens} R.~J.,  {Illingworth} G.,
  {Franx} M.,   {Kriek} M.,  2011, \apjl, 735, L34

\bibitem[\protect\citeauthoryear{{Grazian} et~al.,}{{Grazian}
  et~al.}{2015}]{Grazian2015}
{Grazian} A.,  et~al., 2015, \aap, 575, A96

\bibitem[\protect\citeauthoryear{{Haardt} \& {Madau}}{{Haardt} \&
  {Madau}}{1996}]{Haardt1996}
{Haardt} F.,  {Madau} P.,  1996, \mn@doi [\apj] {10.1086/177035}, \href
  {http://ads.ari.uni-heidelberg.de/abs/1996ApJ...461...20H} {461, 20}

\bibitem[\protect\citeauthoryear{{Heger} \& {Woosley}}{{Heger} \&
  {Woosley}}{2002}]{heger2002}
{Heger} A.,  {Woosley} S.~E.,  2002, \mn@doi [\apj] {10.1086/338487}, \href
  {http://ads.ari.uni-heidelberg.de/abs/2002ApJ...567..532H} {567, 532}

\bibitem[\protect\citeauthoryear{{Hirashita}, {Ferrara}, {Dayal}  \&
  {Ouchi}}{{Hirashita} et~al.}{2014}]{Hirashita2014a}
{Hirashita} H.,  {Ferrara} A.,  {Dayal} P.,   {Ouchi} M.,  2014, \mn@doi
  [\mnras] {10.1093/mnras/stu1290}, \href
  {http://ads.ari.uni-heidelberg.de/abs/2014MNRAS.443.1704H} {443, 1704}

\bibitem[\protect\citeauthoryear{{Howell} et~al.,}{{Howell}
  et~al.}{2010}]{Howell2010}
{Howell} J.~H.,  et~al., 2010, \apj, 715, 572

\bibitem[\protect\citeauthoryear{{Hutter}, {Dayal}, {Partl}  \&
  {M{\"u}ller}}{{Hutter} et~al.}{2014}]{Hutter2014}
{Hutter} A.,  {Dayal} P.,  {Partl} A.~M.,   {M{\"u}ller} V.,  2014, \mn@doi
  [\mnras] {10.1093/mnras/stu791}, \href
  {http://adsabs.harvard.edu/abs/2014MNRAS.441.2861H} {441, 2861}

\bibitem[\protect\citeauthoryear{{Kanekar}, {Wagg}, {Ram Chary}  \&
  {Carilli}}{{Kanekar} et~al.}{2013}]{Kanekar2013}
{Kanekar} N.,  {Wagg} J.,  {Ram Chary} R.,   {Carilli} C.~L.,  2013, \mn@doi
  [\apjl] {10.1088/2041-8205/771/2/L20}, \href
  {http://ads.ari.uni-heidelberg.de/abs/2013ApJ...771L..20K} {771, L20}

\bibitem[\protect\citeauthoryear{{Khakhaleva-Li} \& {Gnedin}}{{Khakhaleva-Li}
  \& {Gnedin}}{2016}]{Khakhaleva2016}
{Khakhaleva-Li} Z.,  {Gnedin} N.~Y.,  2016, \apj, 820, 133

\bibitem[\protect\citeauthoryear{{Knudsen}, {Watson}, {Frayer}, {Christensen},
  {Gallazzi}, {Michalowski}, {Richard}  \& {Zavala}}{{Knudsen}
  et~al.}{2016}]{Knudsen2016}
{Knudsen} K.~K.,  {Watson} D.,  {Frayer} D.,  {Christensen} L.,  {Gallazzi} A.,
   {Michalowski} M.~J.,  {Richard} J.,   {Zavala} J.,  2016, preprint

\bibitem[\protect\citeauthoryear{{Komatsu} et~al.,}{{Komatsu}
  et~al.}{2011}]{Komatsu2011}
{Komatsu} E.,  et~al., 2011, \mn@doi [\apjs] {10.1088/0067-0049/192/2/18},
  \href {http://adsabs.harvard.edu/abs/2011ApJS..192...18K} {192, 18}

\bibitem[\protect\citeauthoryear{{Krumholz}, {Dekel}  \& {McKee}}{{Krumholz}
  et~al.}{2012}]{Krumholz2012}
{Krumholz} M.~R.,  {Dekel} A.,   {McKee} C.~F.,  2012, \apj, 745, 69

\bibitem[\protect\citeauthoryear{{Laporte} et~al.,}{{Laporte}
  et~al.}{2015}]{Laporte2015}
{Laporte} N.,  et~al., 2015, \mn@doi [\aap] {10.1051/0004-6361/201425040},
  \href {http://ads.ari.uni-heidelberg.de/abs/2015A%26A...575A..92L} {575, A92}

\bibitem[\protect\citeauthoryear{{Leitherer} et~al.,}{{Leitherer}
  et~al.}{1999}]{Leitherer1999}
{Leitherer} C.,  et~al., 1999, \mn@doi [\apjs] {10.1086/313233}, \href
  {http://ads.ari.uni-heidelberg.de/abs/1999ApJS..123....3L} {123, 3}

\bibitem[\protect\citeauthoryear{{Livermore}, {Finkelstein}  \&
  {Lotz}}{{Livermore} et~al.}{2016}]{Livermore2016}
{Livermore} R.~C.,  {Finkelstein} S.~L.,   {Lotz} J.~M.,  2016, preprint, \href
  {http://adsabs.harvard.edu/abs/2016arXiv160406799L} {} (\mn@eprint {arXiv}
  {1604.06799})

\bibitem[\protect\citeauthoryear{{Maio}, {Dolag}, {Ciardi}  \&
  {Tornatore}}{{Maio} et~al.}{2007}]{maio2007}
{Maio} U.,  {Dolag} K.,  {Ciardi} B.,   {Tornatore} L.,  2007, \mn@doi [\mnras]
  {10.1111/j.1365-2966.2007.12016.x}, \href
  {http://ads.ari.uni-heidelberg.de/abs/2007MNRAS.379..963M} {379, 963}

\bibitem[\protect\citeauthoryear{{Maio}, {Ciardi}, {Dolag}, {Tornatore}  \&
  {Khochfar}}{{Maio} et~al.}{2010}]{maio2010}
{Maio} U.,  {Ciardi} B.,  {Dolag} K.,  {Tornatore} L.,   {Khochfar} S.,  2010,
  \mn@doi [\mnras] {10.1111/j.1365-2966.2010.17003.x}, \href
  {http://adsabs.harvard.edu/abs/2010MNRAS.407.1003M} {407, 1003}

\bibitem[\protect\citeauthoryear{{Maio}, {Khochfar}, {Johnson}  \&
  {Ciardi}}{{Maio} et~al.}{2011}]{Maio2011}
{Maio} U.,  {Khochfar} S.,  {Johnson} J.~L.,   {Ciardi} B.,  2011, \mn@doi
  [\mnras] {10.1111/j.1365-2966.2011.18455.x}, \href
  {http://ads.ari.uni-heidelberg.de/abs/2011MNRAS.414.1145M} {414, 1145}

\bibitem[\protect\citeauthoryear{{Maiolino}, {Schneider}, {Oliva}, {Bianchi},
  {Ferrara}, {Mannucci}, {Pedani}  \& {Roca Sogorb}}{{Maiolino}
  et~al.}{2004}]{Maiolino2004}
{Maiolino} R.,  {Schneider} R.,  {Oliva} E.,  {Bianchi} S.,  {Ferrara} A.,
  {Mannucci} F.,  {Pedani} M.,   {Roca Sogorb} M.,  2004, \nat, 431, 533

\bibitem[\protect\citeauthoryear{{Maiolino} et~al.,}{{Maiolino}
  et~al.}{2015}]{Maiolino2015}
{Maiolino} R.,  et~al., 2015, \mn@doi [\mnras] {10.1093/mnras/stv1194}, \href
  {http://adsabs.harvard.edu/abs/2015MNRAS.452...54M} {452, 54}

\bibitem[\protect\citeauthoryear{{Mancini}, {Schneider}, {Graziani},
  {Valiante}, {Dayal}, {Maio}, {Ciardi}  \& {Hunt}}{{Mancini}
  et~al.}{2015}]{mancini2015}
{Mancini} M.,  {Schneider} R.,  {Graziani} L.,  {Valiante} R.,  {Dayal} P.,
  {Maio} U.,  {Ciardi} B.,   {Hunt} L.~K.,  2015, \mn@doi [\mnras]
  {10.1093/mnrasl/slv070}, \href
  {http://adsabs.harvard.edu/abs/2015MNRAS.451L..70M} {451, L70}

\bibitem[\protect\citeauthoryear{{Marassi}, {Chiaki}, {Schneider}, {Limongi},
  {Omukai}, {Nozawa}, {Chieffi}  \& {Yoshida}}{{Marassi}
  et~al.}{2014}]{Marassi2014}
{Marassi} S.,  {Chiaki} G.,  {Schneider} R.,  {Limongi} M.,  {Omukai} K.,
  {Nozawa} T.,  {Chieffi} A.,   {Yoshida} N.,  2014, \mn@doi [\apj]
  {10.1088/0004-637X/794/2/100}, \href
  {http://adsabs.harvard.edu/abs/2014ApJ...794..100M} {794, 100}

\bibitem[\protect\citeauthoryear{{Marassi}, {Schneider}, {Limongi}, {Chieffi},
  {Bocchio}  \& {Bianchi}}{{Marassi} et~al.}{2015}]{Marassi2015}
{Marassi} S.,  {Schneider} R.,  {Limongi} M.,  {Chieffi} A.,  {Bocchio} M.,
  {Bianchi} S.,  2015, \mn@doi [\mnras] {10.1093/mnras/stv2267}, \href
  {http://adsabs.harvard.edu/abs/2015MNRAS.454.4250M} {454, 4250}

\bibitem[\protect\citeauthoryear{{McKee}}{{McKee}}{1989}]{McKee1989}
{McKee} C.,  1989, in {Allamandola} L.~J.,  {Tielens} A.~G.~G.~M.,  eds,  IAU
  Symposium Vol. 135, Interstellar Dust. p.~431

\bibitem[\protect\citeauthoryear{{McLeod}, {McLure}, {Dunlop}, {Robertson},
  {Ellis}  \& {Targett}}{{McLeod} et~al.}{2015}]{McLeod2015}
{McLeod} D.~J.,  {McLure} R.~J.,  {Dunlop} J.~S.,  {Robertson} B.~E.,  {Ellis}
  R.~S.,   {Targett} T.~A.,  2015, \mnras, 450, 3032

\bibitem[\protect\citeauthoryear{{McLeod}, {McLure}  \& {Dunlop}}{{McLeod}
  et~al.}{2016}]{McLeod2016}
{McLeod} D.~J.,  {McLure} R.~J.,   {Dunlop} J.~S.,  2016, preprint, 1602.05199

\bibitem[\protect\citeauthoryear{{McLure}, {Cirasuolo}, {Dunlop}, {Foucaud}  \&
  {Almaini}}{{McLure} et~al.}{2009}]{mclure2009}
{McLure} R.~J.,  {Cirasuolo} M.,  {Dunlop} J.~S.,  {Foucaud} S.,   {Almaini}
  O.,  2009, \mn@doi [\mnras] {10.1111/j.1365-2966.2009.14677.x}, \href
  {http://adsabs.harvard.edu/abs/2009MNRAS.395.2196M} {395, 2196}

\bibitem[\protect\citeauthoryear{{McLure}, {Dunlop}, {Cirasuolo}, {Koekemoer},
  {Sabbi}, {Stark}, {Targett}  \& {Ellis}}{{McLure} et~al.}{2010}]{mclure2010}
{McLure} R.~J.,  {Dunlop} J.~S.,  {Cirasuolo} M.,  {Koekemoer} A.~M.,  {Sabbi}
  E.,  {Stark} D.~P.,  {Targett} T.~A.,   {Ellis} R.~S.,  2010, \mn@doi
  [\mnras] {10.1111/j.1365-2966.2009.16176.x}, \href
  {http://adsabs.harvard.edu/abs/2010MNRAS.403..960M} {403, 960}

\bibitem[\protect\citeauthoryear{{McLure} et~al.,}{{McLure}
  et~al.}{2013}]{McLure2013}
{McLure} R.~J.,  et~al., 2013, \mnras, 432, 2696

\bibitem[\protect\citeauthoryear{{Meurer}, {Heckman}  \& {Calzetti}}{{Meurer}
  et~al.}{1999}]{meurer1999}
{Meurer} G.~R.,  {Heckman} T.~M.,   {Calzetti} D.,  1999, \mn@doi [\apj]
  {10.1086/307523}, \href {http://adsabs.harvard.edu/abs/1999ApJ...521...64M}
  {521, 64}

\bibitem[\protect\citeauthoryear{{Micha{\l}owski}}{{Micha{\l}owski}}{2015}]{Michalowski2015}
{Micha{\l}owski} M.~J.,  2015, \aap, 577, A80

\bibitem[\protect\citeauthoryear{{Murray}}{{Murray}}{2011}]{Murray2011}
{Murray} N.,  2011, \apj, 729, 133

\bibitem[\protect\citeauthoryear{{Nanni}, {Bressan}, {Marigo}  \&
  {Girardi}}{{Nanni} et~al.}{2013}]{Nanni2013}
{Nanni} A.,  {Bressan} A.,  {Marigo} P.,   {Girardi} L.,  2013, \mn@doi
  [\mnras] {10.1093/mnras/stt1175}, \href
  {http://adsabs.harvard.edu/abs/2013MNRAS.434.2390N} {434, 2390}

\bibitem[\protect\citeauthoryear{{Nozawa}, {Kozasa}, {Umeda}, {Maeda}  \&
  {Nomoto}}{{Nozawa} et~al.}{2003}]{Nozawa2003}
{Nozawa} T.,  {Kozasa} T.,  {Umeda} H.,  {Maeda} K.,   {Nomoto} K.,  2003,
  \mn@doi [\apj] {10.1086/379011}, \href
  {http://adsabs.harvard.edu/abs/2003ApJ...598..785N} {598, 785}

\bibitem[\protect\citeauthoryear{{Nozawa}, {Kozasa}  \& {Habe}}{{Nozawa}
  et~al.}{2006}]{Nozawa2006}
{Nozawa} T.,  {Kozasa} T.,   {Habe} A.,  2006, \apj, 648, 435

\bibitem[\protect\citeauthoryear{{Oesch} et~al.,}{{Oesch}
  et~al.}{2010}]{oesch2010}
{Oesch} P.~A.,  et~al., 2010, \mn@doi [\apjl] {10.1088/2041-8205/709/1/L16},
  \href {http://adsabs.harvard.edu/abs/2010ApJ...709L..16O} {709, L16}

\bibitem[\protect\citeauthoryear{{Oesch} et~al.,}{{Oesch}
  et~al.}{2013}]{Oesch2013}
{Oesch} P.~A.,  et~al., 2013, \mn@doi [\apj] {10.1088/0004-637X/772/2/136},
  \href {http://adsabs.harvard.edu/abs/2013ApJ...772..136O} {772, 136}

\bibitem[\protect\citeauthoryear{{Oesch} et~al.,}{{Oesch}
  et~al.}{2014}]{Oesch2014}
{Oesch} P.~A.,  et~al., 2014, \apj, 786, 108

\bibitem[\protect\citeauthoryear{{Oesch} et~al.,}{{Oesch}
  et~al.}{2016}]{Oesch2016}
{Oesch} P.~A.,  et~al., 2016, preprint, 1603.00461

\bibitem[\protect\citeauthoryear{{Ota} et~al.,}{{Ota} et~al.}{2014}]{Ota2014}
{Ota} K.,  et~al., 2014, \mn@doi [\apj] {10.1088/0004-637X/792/1/34}, \href
  {http://ads.ari.uni-heidelberg.de/abs/2014ApJ...792...34O} {792, 34}

\bibitem[\protect\citeauthoryear{{Ouchi} et~al.,}{{Ouchi}
  et~al.}{2013}]{Ouchi2013}
{Ouchi} M.,  et~al., 2013, \mn@doi [\apj] {10.1088/0004-637X/778/2/102}, \href
  {http://ads.ari.uni-heidelberg.de/abs/2013ApJ...778..102O} {778, 102}

\bibitem[\protect\citeauthoryear{{Overzier} et~al.,}{{Overzier}
  et~al.}{2011}]{Overzier2011}
{Overzier} R.~A.,  et~al., 2011, \apjl, 726, L7

\bibitem[\protect\citeauthoryear{{Padovani} \& {Matteucci}}{{Padovani} \&
  {Matteucci}}{1993}]{padovani1993}
{Padovani} P.,  {Matteucci} F.,  1993, \mn@doi [\apj] {10.1086/173212}, \href
  {http://adsabs.harvard.edu/abs/1993ApJ...416...26P} {416, 26}

\bibitem[\protect\citeauthoryear{{Pei}}{{Pei}}{1992}]{Pei1992}
{Pei} Y.~C.,  1992, \mn@doi [\apj] {10.1086/171637}, \href
  {http://ads.ari.uni-heidelberg.de/abs/1992ApJ...395..130P} {395, 130}

\bibitem[\protect\citeauthoryear{{Perley} et~al.,}{{Perley}
  et~al.}{2010}]{Perley2010}
{Perley} D.~A.,  et~al., 2010, \mnras, 406, 2473

\bibitem[\protect\citeauthoryear{{Pettini}, {Kellogg}, {Steidel}, {Dickinson},
  {Adelberger}  \& {Giavalisco}}{{Pettini} et~al.}{1998}]{Pettini1998}
{Pettini} M.,  {Kellogg} M.,  {Steidel} C.~C.,  {Dickinson} M.,  {Adelberger}
  K.~L.,   {Giavalisco} M.,  1998, \apj, 508, 539

\bibitem[\protect\citeauthoryear{{Reddy}, {Erb}, {Pettini}, {Steidel}  \&
  {Shapley}}{{Reddy} et~al.}{2010}]{Reddy2010}
{Reddy} N.~A.,  {Erb} D.~K.,  {Pettini} M.,  {Steidel} C.~C.,   {Shapley}
  A.~E.,  2010, \apj, 712, 1070

\bibitem[\protect\citeauthoryear{{Rogers} et~al.,}{{Rogers}
  et~al.}{2014}]{Rogers2014}
{Rogers} A.~B.,  et~al., 2014, \mn@doi [\mnras] {10.1093/mnras/stu558}, \href
  {http://adsabs.harvard.edu/abs/2014MNRAS.440.3714R} {440, 3714}

\bibitem[\protect\citeauthoryear{{Salvaterra}, {Ferrara}  \&
  {Dayal}}{{Salvaterra} et~al.}{2011}]{Salvaterra2011}
{Salvaterra} R.,  {Ferrara} A.,   {Dayal} P.,  2011, \mn@doi [\mnras]
  {10.1111/j.1365-2966.2010.18155.x}, \href
  {http://adsabs.harvard.edu/abs/2011MNRAS.414..847S} {414, 847}

\bibitem[\protect\citeauthoryear{{Salvaterra}, {Maio}, {Ciardi}  \&
  {Campisi}}{{Salvaterra} et~al.}{2013}]{salvaterra2013}
{Salvaterra} R.,  {Maio} U.,  {Ciardi} B.,   {Campisi} M.~A.,  2013, \mnras,
  429, 2718

\bibitem[\protect\citeauthoryear{{Sarangi} \& {Cherchneff}}{{Sarangi} \&
  {Cherchneff}}{2013}]{Sarangi2013}
{Sarangi} A.,  {Cherchneff} I.,  2013, \mn@doi [\apj]
  {10.1088/0004-637X/776/2/107}, \href
  {http://adsabs.harvard.edu/abs/2013ApJ...776..107S} {776, 107}

\bibitem[\protect\citeauthoryear{{Schaerer}, {Nakajima}, {Dessauges-Zavadsky},
  {Walth}, {Rujopakarn}, {Richard}  \& {Egami}}{{Schaerer}
  et~al.}{2015}]{Schaerer2015}
{Schaerer} D.,  {Nakajima} K.,  {Dessauges-Zavadsky} M.,  {Walth} G.,
  {Rujopakarn} W.,  {Richard} J.,   {Egami} E.,  2015, IAU General Assembly,
  \href {http://ads.ari.uni-heidelberg.de/abs/2015IAUGA..2258419S} {22, 58419}

\bibitem[\protect\citeauthoryear{{Schneider}, {Ferrara}  \&
  {Salvaterra}}{{Schneider} et~al.}{2004}]{Schneider2004}
{Schneider} R.,  {Ferrara} A.,   {Salvaterra} R.,  2004, \mn@doi [\mnras]
  {10.1111/j.1365-2966.2004.07876.x}, \href
  {http://adsabs.harvard.edu/abs/2004MNRAS.351.1379S} {351, 1379}

\bibitem[\protect\citeauthoryear{{Schneider}, {Valiante}, {Ventura},
  {dell'Agli}, {Di Criscienzo}, {Hirashita}  \& {Kemper}}{{Schneider}
  et~al.}{2014}]{Schneider2014}
{Schneider} R.,  {Valiante} R.,  {Ventura} P.,  {dell'Agli} F.,  {Di
  Criscienzo} M.,  {Hirashita} H.,   {Kemper} F.,  2014, \mn@doi [\mnras]
  {10.1093/mnras/stu861}, \href
  {http://adsabs.harvard.edu/abs/2014MNRAS.442.1440S} {442, 1440}

\bibitem[\protect\citeauthoryear{{Schneider}, {Hunt}  \&
  {Valiante}}{{Schneider} et~al.}{2016}]{Schneider2016}
{Schneider} R.,  {Hunt} L.,   {Valiante} R.,  2016, \mnras, 457, 1842

\bibitem[\protect\citeauthoryear{{Shimizu}, {Inoue}, {Okamoto}  \&
  {Yoshida}}{{Shimizu} et~al.}{2014}]{Shimizu2014}
{Shimizu} I.,  {Inoue} A.~K.,  {Okamoto} T.,   {Yoshida} N.,  2014, \mn@doi
  [\mnras] {10.1093/mnras/stu265}, \href
  {http://adsabs.harvard.edu/abs/2014MNRAS.440..731S} {440, 731}

\bibitem[\protect\citeauthoryear{{Siana} et~al.,}{{Siana}
  et~al.}{2009}]{Siana2009}
{Siana} B.,  et~al., 2009, \apj, 698, 1273

\bibitem[\protect\citeauthoryear{{Song} et~al.,}{{Song}
  et~al.}{2016}]{Song2016}
{Song} M.,  et~al., 2016, \mn@doi [\apj] {10.3847/0004-637X/825/1/5}, \href
  {http://adsabs.harvard.edu/abs/2016ApJ...825....5S} {825, 5}

\bibitem[\protect\citeauthoryear{{Springel}}{{Springel}}{2005}]{springel2005}
{Springel} V.,  2005, \mn@doi [\mnras] {10.1111/j.1365-2966.2005.09655.x},
  \href {http://adsabs.harvard.edu/abs/2005MNRAS.364.1105S} {364, 1105}

\bibitem[\protect\citeauthoryear{{Springel} \& {Hernquist}}{{Springel} \&
  {Hernquist}}{2003}]{Springel2003}
{Springel} V.,  {Hernquist} L.,  2003, \mn@doi [\mnras]
  {10.1046/j.1365-8711.2003.06206.x}, \href
  {http://ads.ari.uni-heidelberg.de/abs/2003MNRAS.339..289S} {339, 289}

\bibitem[\protect\citeauthoryear{{Stanway}, {McMahon}  \& {Bunker}}{{Stanway}
  et~al.}{2005}]{Stanway2005}
{Stanway} E.~R.,  {McMahon} R.~G.,   {Bunker} A.~J.,  2005, \mn@doi [\mnras]
  {10.1111/j.1365-2966.2005.08977.x}, \href
  {http://adsabs.harvard.edu/abs/2005MNRAS.359.1184S} {359, 1184}

\bibitem[\protect\citeauthoryear{{Stratta}, {Maiolino}, {Fiore}  \&
  {D'Elia}}{{Stratta} et~al.}{2007}]{Stratta2007}
{Stratta} G.,  {Maiolino} R.,  {Fiore} F.,   {D'Elia} V.,  2007, \apjl, 661, L9

\bibitem[\protect\citeauthoryear{{Takeuchi}, {Buat}, {Heinis}, {Giovannoli},
  {Yuan}, {Iglesias-P{\'a}ramo}, {Murata}  \& {Burgarella}}{{Takeuchi}
  et~al.}{2010}]{Takeuchi2010}
{Takeuchi} T.~T.,  {Buat} V.,  {Heinis} S.,  {Giovannoli} E.,  {Yuan} F.-T.,
  {Iglesias-P{\'a}ramo} J.,  {Murata} K.~L.,   {Burgarella} D.,  2010, \aap,
  514, A4

\bibitem[\protect\citeauthoryear{{Talia} et~al.,}{{Talia}
  et~al.}{2015}]{Talia2015}
{Talia} M.,  et~al., 2015, \aap, 582, A80

\bibitem[\protect\citeauthoryear{{Thielemann} et~al.,}{{Thielemann}
  et~al.}{2003}]{thielemann2003}
{Thielemann} F.-K.,  et~al., 2003, \mn@doi [Nuclear Physics A]
  {10.1016/S0375-9474(03)00704-8}, \href
  {http://adsabs.harvard.edu/abs/2003NuPhA.718..139T} {718, 139}

\bibitem[\protect\citeauthoryear{{Tilvi} et~al.,}{{Tilvi}
  et~al.}{2013}]{Tilvi2013}
{Tilvi} V.,  et~al., 2013, \mn@doi [\apj] {10.1088/0004-637X/768/1/56}, \href
  {http://adsabs.harvard.edu/abs/2013ApJ...768...56T} {768, 56}

\bibitem[\protect\citeauthoryear{{Todini} \& {Ferrara}}{{Todini} \&
  {Ferrara}}{2001}]{Todini2001}
{Todini} P.,  {Ferrara} A.,  2001, \mn@doi [\mnras]
  {10.1046/j.1365-8711.2001.04486.x}, \href
  {http://adsabs.harvard.edu/abs/2001MNRAS.325..726T} {325, 726}

\bibitem[\protect\citeauthoryear{{Tornatore}, {Borgani}, {Dolag}  \&
  {Matteucci}}{{Tornatore} et~al.}{2007}]{tornatore2007}
{Tornatore} L.,  {Borgani} S.,  {Dolag} K.,   {Matteucci} F.,  2007, \mn@doi
  [\mnras] {10.1111/j.1365-2966.2007.12070.x}, \href
  {http://adsabs.harvard.edu/abs/2007MNRAS.382.1050T} {382, 1050}

\bibitem[\protect\citeauthoryear{{Valiante}, {Schneider}, {Bianchi}  \&
  {Andersen}}{{Valiante} et~al.}{2009}]{Valiante2009}
{Valiante} R.,  {Schneider} R.,  {Bianchi} S.,   {Andersen} A.~C.,  2009,
  \mn@doi [\mnras] {10.1111/j.1365-2966.2009.15076.x}, \href
  {http://adsabs.harvard.edu/abs/2009MNRAS.397.1661V} {397, 1661}

\bibitem[\protect\citeauthoryear{{Valiante}, {Schneider}, {Salvadori}  \&
  {Bianchi}}{{Valiante} et~al.}{2011}]{Valiante2011}
{Valiante} R.,  {Schneider} R.,  {Salvadori} S.,   {Bianchi} S.,  2011, \mn@doi
  [\mnras] {10.1111/j.1365-2966.2011.19168.x}, \href
  {http://ads.ari.uni-heidelberg.de/abs/2011MNRAS.416.1916V} {416, 1916}

\bibitem[\protect\citeauthoryear{{V{\'a}zquez} \& {Leitherer}}{{V{\'a}zquez} \&
  {Leitherer}}{2005}]{Vazquez2005}
{V{\'a}zquez} G.~A.,  {Leitherer} C.,  2005, \mn@doi [\apj] {10.1086/427866},
  \href {http://ads.ari.uni-heidelberg.de/abs/2005ApJ...621..695V} {621, 695}

\bibitem[\protect\citeauthoryear{{Ventura} et~al.,}{{Ventura}
  et~al.}{2012a}]{Ventura2012a}
{Ventura} P.,  et~al., 2012a, \mn@doi [\mnras]
  {10.1111/j.1365-2966.2011.20129.x}, \href
  {http://adsabs.harvard.edu/abs/2012MNRAS.420.1442V} {420, 1442}

\bibitem[\protect\citeauthoryear{{Ventura} et~al.,}{{Ventura}
  et~al.}{2012b}]{Ventura2012b}
{Ventura} P.,  et~al., 2012b, \mn@doi [\mnras]
  {10.1111/j.1365-2966.2012.21403.x}, \href
  {http://adsabs.harvard.edu/abs/2012MNRAS.424.2345V} {424, 2345}

\bibitem[\protect\citeauthoryear{{Ventura}, {Dell'Agli}, {Schneider}, {Di
  Criscienzo}, {Rossi}, {La Franca}, {Gallerani}  \& {Valiante}}{{Ventura}
  et~al.}{2014}]{Ventura2014}
{Ventura} P.,  {Dell'Agli} F.,  {Schneider} R.,  {Di Criscienzo} M.,  {Rossi}
  C.,  {La Franca} F.,  {Gallerani} S.,   {Valiante} R.,  2014, \mn@doi
  [\mnras] {10.1093/mnras/stu028}, \href
  {http://adsabs.harvard.edu/abs/2014MNRAS.439..977V} {439, 977}

\bibitem[\protect\citeauthoryear{{Waters}, {Wilkins}, {Di Matteo}, {Feng},
  {Croft}  \& {Nagai}}{{Waters} et~al.}{2016}]{Waters2016}
{Waters} D.,  {Wilkins} S.,  {Di Matteo} T.,  {Feng} Y.,  {Croft} R.,   {Nagai}
  D.,  2016, preprint

\bibitem[\protect\citeauthoryear{{Watson}, {Christensen}, {Knudsen}, {Richard},
  {Gallazzi}  \& {Micha{\l}owski}}{{Watson} et~al.}{2015}]{Watson2015}
{Watson} D.,  {Christensen} L.,  {Knudsen} K.~K.,  {Richard} J.,  {Gallazzi}
  A.,   {Micha{\l}owski} M.~J.,  2015, \mn@doi [\nat] {10.1038/nature14164},
  \href {http://ads.ari.uni-heidelberg.de/abs/2015Natur.519..327W} {519, 327}

\bibitem[\protect\citeauthoryear{{Weingartner} \& {Draine}}{{Weingartner} \&
  {Draine}}{2001}]{Weingartner2001}
{Weingartner} J.~C.,  {Draine} B.~T.,  2001, \mn@doi [\apj] {10.1086/318651},
  \href {http://ads.ari.uni-heidelberg.de/abs/2001ApJ...548..296W} {548, 296}

\bibitem[\protect\citeauthoryear{{Wilkins}, {Bunker}, {Stanway}, {Lorenzoni}
  \& {Caruana}}{{Wilkins} et~al.}{2011}]{Wilkins2011}
{Wilkins} S.~M.,  {Bunker} A.~J.,  {Stanway} E.,  {Lorenzoni} S.,   {Caruana}
  J.,  2011, \mn@doi [\mnras] {10.1111/j.1365-2966.2011.19315.x}, \href
  {http://adsabs.harvard.edu/abs/2011MNRAS.417..717W} {417, 717}

\bibitem[\protect\citeauthoryear{{Wilkins}, {Gonzalez-Perez}, {Lacey}  \&
  {Baugh}}{{Wilkins} et~al.}{2012}]{Wilkins2012b}
{Wilkins} S.~M.,  {Gonzalez-Perez} V.,  {Lacey} C.~G.,   {Baugh} C.~M.,  2012,
  \mn@doi [\mnras] {10.1111/j.1365-2966.2012.21344.x}, \href
  {http://adsabs.harvard.edu/abs/2012MNRAS.424.1522W} {424, 1522}

\bibitem[\protect\citeauthoryear{{Wilkins}, {Bunker}, {Coulton}, {Croft},
  {Matteo}, {Khandai}  \& {Feng}}{{Wilkins} et~al.}{2013}]{Wilkins2013}
{Wilkins} S.~M.,  {Bunker} A.,  {Coulton} W.,  {Croft} R.,  {Matteo} T.~D.,
  {Khandai} N.,   {Feng} Y.,  2013, \mn@doi [\mnras] {10.1093/mnras/stt096},
  \href {http://adsabs.harvard.edu/abs/2013MNRAS.430.2885W} {430, 2885}

\bibitem[\protect\citeauthoryear{{Woosley} \& {Weaver}}{{Woosley} \&
  {Weaver}}{1995}]{woosley1995}
{Woosley} S.~E.,  {Weaver} T.~A.,  1995, \mn@doi [\apjs] {10.1086/192237},
  \href {http://adsabs.harvard.edu/abs/1995ApJS..101..181W} {101, 181}

\bibitem[\protect\citeauthoryear{{Yoshida}, {Abel}, {Hernquist}  \&
  {Sugiyama}}{{Yoshida} et~al.}{2003}]{yoshida2003}
{Yoshida} N.,  {Abel} T.,  {Hernquist} L.,   {Sugiyama} N.,  2003, \mn@doi
  [\apj] {10.1086/375810}, \href
  {http://adsabs.harvard.edu/abs/2003ApJ...592..645Y} {592, 645}

\bibitem[\protect\citeauthoryear{{Zafar}, {Watson}, {Malesani}, {Vreeswijk},
  {Fynbo}, {Hjorth}, {Levan}  \& {Micha{\l}owski}}{{Zafar}
  et~al.}{2010}]{Zafar2010}
{Zafar} T.,  {Watson} D.~J.,  {Malesani} D.,  {Vreeswijk} P.~M.,  {Fynbo}
  J.~P.~U.,  {Hjorth} J.,  {Levan} A.~J.,   {Micha{\l}owski} M.~J.,  2010,
  \aap, 515

\bibitem[\protect\citeauthoryear{{Zavala} et~al.,}{{Zavala}
  et~al.}{2015}]{Zavala2015}
{Zavala} J.~A.,  et~al., 2015, \mnras, 453, L88

\bibitem[\protect\citeauthoryear{{Zhukovska}, {Gail}  \&
  {Trieloff}}{{Zhukovska} et~al.}{2008}]{Zhukovska2008}
{Zhukovska} S.,  {Gail} H.-P.,   {Trieloff} M.,  2008, \mn@doi [\aap]
  {10.1051/0004-6361:20077789}, \href
  {http://ads.ari.uni-heidelberg.de/abs/2008A%26A...479..453Z} {479, 453}

\bibitem[\protect\citeauthoryear{{Zitrin} et~al.,}{{Zitrin}
  et~al.}{2015}]{Zitrin2015}
{Zitrin} A.,  et~al., 2015, \mn@doi [\apjl] {10.1088/2041-8205/810/1/L12},
  \href {http://adsabs.harvard.edu/abs/2015ApJ...810L..12Z} {810, L12}

\bibitem[\protect\citeauthoryear{{de Bennassuti}, {Schneider}, {Valiante}  \&
  {Salvadori}}{{de Bennassuti} et~al.}{2014}]{deBennassuti2014}
{de Bennassuti} M.,  {Schneider} R.,  {Valiante} R.,   {Salvadori} S.,  2014,
  \mn@doi [\mnras] {10.1093/mnras/stu1962}, \href
  {http://ads.ari.uni-heidelberg.de/abs/2014MNRAS.445.3039D} {445, 3039}

\bibitem[\protect\citeauthoryear{{van den Hoek} \& {Groenewegen}}{{van den
  Hoek} \& {Groenewegen}}{1997}]{vandenhoek1997}
{van den Hoek} L.~B.,  {Groenewegen} M.~A.~T.,  1997, \mn@doi [\aaps]
  {10.1051/aas:1997162}, \href
  {http://adsabs.harvard.edu/abs/1997A%26AS..123..305V} {123, 305}

\makeatother
\end{thebibliography}
\label{lastpage}

\end{document}